\documentclass[trsc,nonblindrev]{template}
\OneAndAHalfSpacedXI
\usepackage{natbib}
\usepackage{dsfont}
 \bibpunct[, ]{(}{)}{,}{a}{}{,}%
 \def\bibfont{\fontsize{11}{13.5}\selectfont}%

\TheoremsNumberedThrough
\EquationsNumberedThrough

\usepackage{booktabs}
 \usepackage{bm}
\usepackage[short]{ optidef }

\usepackage[utf8]{inputenc}
\usepackage[margin=1in]{geometry} 
\usepackage{enumitem}
\usepackage{graphicx}
\usepackage{lipsum}
\usepackage{caption}
\usepackage{subcaption}
\usepackage{soul}
\usepackage[colorlinks=true,breaklinks=true,bookmarks=true,urlcolor=blue,
     citecolor=blue,linkcolor=blue,bookmarksopen=false,draft=false]{hyperref}
\usepackage[dvipsnames,table,xcdraw]{xcolor}
\usepackage{siunitx}
\usepackage{multirow}
\usepackage{hypcap}
\usepackage{float}
\usepackage{cleveref}
\usepackage{tikz}
\usetikzlibrary{shapes.misc}
\usetikzlibrary{shapes.geometric}
\tikzset{cross/.style={cross out, draw=black, minimum size=20*(#1-\pgflinewidth), inner sep=0pt, outer sep=0pt},
cross/.default={1pt}}
\usepackage{pgfplots}
\usepackage{stmaryrd}

\usepackage[ruled,vlined,linesnumbered]{algorithm2e}

\usepackage[symbol]{footmisc}

\newcommand{\type}[2]{\tau_{#1,#2}}
\newcommand{\front}[3]{\widetilde{\O}^{#1}_{#2,#3}}

\def\X{\mathcal{X}}
\def\O{\mathcal{O}}
\def\S{\mathcal{S}}
\def\A{\mathcal{A}}
\def\C{\mathcal{C}}

\usetikzlibrary{shapes.geometric}
\usetikzlibrary{backgrounds}
\usetikzlibrary{positioning}
\usetikzlibrary{patterns}
\usetikzlibrary{arrows,decorations.markings}
\definecolor{darkred} {HTML}{b34700}

\defcitealias{hqda21}{HQDA 2021}
\defcitealias{jp3-0}{JCS 2017}
\defcitealias{jp5-0}{JCS 2020}
\defcitealias{ar71-32}{HQDA 2019}
\defcitealias{jp1-02}{JCS 2016}
\defcitealias{fm6-0}{HQDA 2016}

\let\ForAll\forall
\renewcommand\forall{\ForAll\;}

\let\Exists\exists
\renewcommand\exists{\Exists\;}

\makeatletter
\newcommand{\settitle}{\@maketitle}
\makeatother

\begin{document}
\TITLE{Dynamic Operational Planning in Warfare: A Stochastic Game Approach to Military Campaigns}
\ARTICLEAUTHORS{%
\AUTHOR{Joseph E. McCarthy, Mathieu Dahan, Chelsea C. White III}
\AFF{H. Milton Stewart School of Industrial and Systems Engineering, Georgia Institute of Technology, Atlanta, Georgia \\
\{\EMAIL{jmccarthy@gatech.edu}, \EMAIL{mathieu.dahan@isye.gatech.edu}, \EMAIL{cw196@gatech.edu}\}}
} 

\RUNTITLE{Stochastic Game for Dynamic Operational Planning in Military Campaigns}
\RUNAUTHOR{McCarthy, Dahan, and White}

\ABSTRACT{We study a two-player discounted zero-sum stochastic game model for dynamic operational planning in military campaigns. At each stage, the players manage multiple commanders who order military actions on objectives that have an open line of control. When a battle over the control of an objective occurs, its stochastic outcome depends on the actions and the enabling support provided by the control of other objectives. Each player aims to maximize the cumulative number of objectives they control, weighted by their criticality. To solve this large-scale stochastic game, we derive properties of its Markov perfect equilibria by leveraging the logistics and military operational command and control structure. We show the consequential isotonicity of the optimal value function with respect to the partially ordered state space, which in turn leads to a significant reduction of the state and action spaces. We also accelerate Shapley’s value iteration algorithm by eliminating dominated actions and investigating pure equilibria of the matrix game solved at each iteration. We demonstrate the computational value of our equilibrium results on a case study that reflects representative operational-level military campaigns with geopolitical implications. Our analysis reveals a complex interplay between the game’s parameters and dynamics in equilibrium, resulting in new military insights for campaign analysts. 
%Our work provides a novel approach for military campaign analysts to model dynamic, operational-level warfare.
}


\HISTORY{This article was first submitted on March 1, 2024.}

\maketitle

%!TEX root = NRL Manuscript.tex
\section{Introduction}

\subsection{Motivation}

Military leadership plays an indispensable role in a nation state's security in times of great competition. Ideally, the leadership can continue to deter adversaries from escalating to kinetic warfare \citepalias{hqda21}; however, planning for armed conflict is essential. Rising geopolitical unrest demonstrates the increasing likelihood of kinetic warfare between great powers \citep{gara22}. This is evidenced by the continued Russo-Ukrainian war that escalated to overt, armed conflict in February 2022. This global context demands that senior military leadership and their staffs continue to conduct planning that unifies the tactical, operational, and strategic levels to support national security objectives \citepalias{jp5-0}. 

Operational-level warfare links the tactical employment of forces to national strategic objectives \citepalias{jp3-0}. At this level, Joint Force Commanders lead component commanders (e.g., air, land, and maritime) to fight conflicts. A sequence of operations and battles form a military \emph{campaign} \citep{lynes14}, which strategists, planners, and analysts alike analyze to recommend operational plans and yield geopolitical insights for senior military leadership \citep{muel16, shla16, flan19, maza19}.

However, the main challenge in analyzing military campaigns comes from their intrinsic uncertainty \citep{teco21} that extends from three sources: the adversary's plan, the interconnected nature of military operations, and the dynamic flow of warfare. Because the adversary's plan is unknown, amidst a complex operational environment \citepalias{jp5-0}, a battle's outcome is uncertain. Synchronizing operations across interconnected commanders demands joint planning, communication, and coordination to align efforts \citepalias{jp3-0}. Finally, the dynamic flow of a military campaign leads to uncertainty in the transition between potential battles. For instance during World War II, a successful allied D-Day invasion eventually led to Operation Market Garden, where U.S. and U.K. setbacks resulted in Soviet forces reaching Berlin first \citep{hist19}. This dynamic facet requires that future uncertain outcomes must be considered to optimize present decisions.

Existing methods for campaign analysis include wargaming and combat simulation \citep{turn22}. While effective, these tools do not account for the uncertain or behavioral dynamics of military campaigns, and take substantial time and resources. Furthermore, game theoretic models that allocate resources for the control of military objectives do not consider pivotal campaign aspects including the dynamic, sequenced nature of military battles, supply chain requirements, or military command structure \citep{wash14}. Seeking to augment current methods with a faster technique that scales to evaluate many inputs, prompts our research question: \emph{How may we design dynamic military operational plans and yield timely assessments and insights for senior leadership?}

\subsection{Contributions}

To address the research question, we propose a novel two-player, discounted, zero-sum, stochastic game model for dynamic operational planning in military campaigns, building upon the static game model of \cite{haywood54}. The features of the model account for key military characteristics such as the coordination of multiple commanders, the need for established supply lines, and the stochasticity of battle outcomes that depend on the control of nearby objectives.

By leveraging the logistics and military operational command and control structure, we derive properties satisfied by the Markov perfect equilibria of the game. Under practically motivated assumptions, we show the consequential isotonicity of the optimal value function with respect to the partially ordered state space  (Theorem \ref{thm:isotone}). This main result, along with game-theoretic arguments, permit us to determine the set of achievable states as well as properties of policy profiles in equilibrium (Proposition \ref{prop:reduce}). These properties lead to a significant reduction of the state and action spaces, which in turn enables the use of Shapley's value iteration algorithm to solve this large-scale game.

In the special case of a campaign with one commander, we further show that the matrix game solved at most states within the value iteration algorithm admits weakly dominated strategies, and even admits a pure equilibrium when the commander manages a single axis of objectives (Proposition \ref{prop:reduction}). These structural results lead us to design an accelerated value iteration algorithm (Algorithms \ref{alg:AZS}-\ref{alg:AVI}) that searches for pure equilibria or eliminates weakly dominated actions before solving the matrix game using linear programming.

We then design a representative case study, built upon a fictional geopolitical scenario. We analyze and compare the players' mixed strategies at different states in equilibrium, and highlight a complex behavior that depends on  the objective criticality, the probabilistic interdependencies between objectives, and the dynamics of the game. We also show that strategic investment decisions must be carefully timed, as they have varied impacts on the optimal value of the game at different initial states. Finally, our equilibrium results permit us to solve the stochastic game for all considered military campaigns, using our accelerated value iteration algorithm that achieves a 72\% runtime reduction in comparison with the classical value iteration algorithm. Our analysis leads to novel operational insights that can be exploited by military leadership.

The remainder of this article is organized as follows: Section \ref{sec:lit} briefly discusses military campaign analysis and current tools. The review then covers existing stochastic game literature pertinent to our military domain. We formulate the stochastic game in Section \ref{sec:Pb}. We then derive our equilibrium results and present our accelerated value iteration algorithm in Section \ref{sec:approaches}. In Section \ref{sec:case_study}, we present our computational results and military insights from our case study. Section \ref{sec:conclusion} provides concluding remarks and avenues for future research. Finally, the mathematical proofs of our results are derived in Appendix \ref{app:deriv}.

%!TEX root = NRL Manuscript.tex
\section{Literature Review} \label{sec:lit}

\begin{figure}[H] 
\centering
\includegraphics[scale=.25]{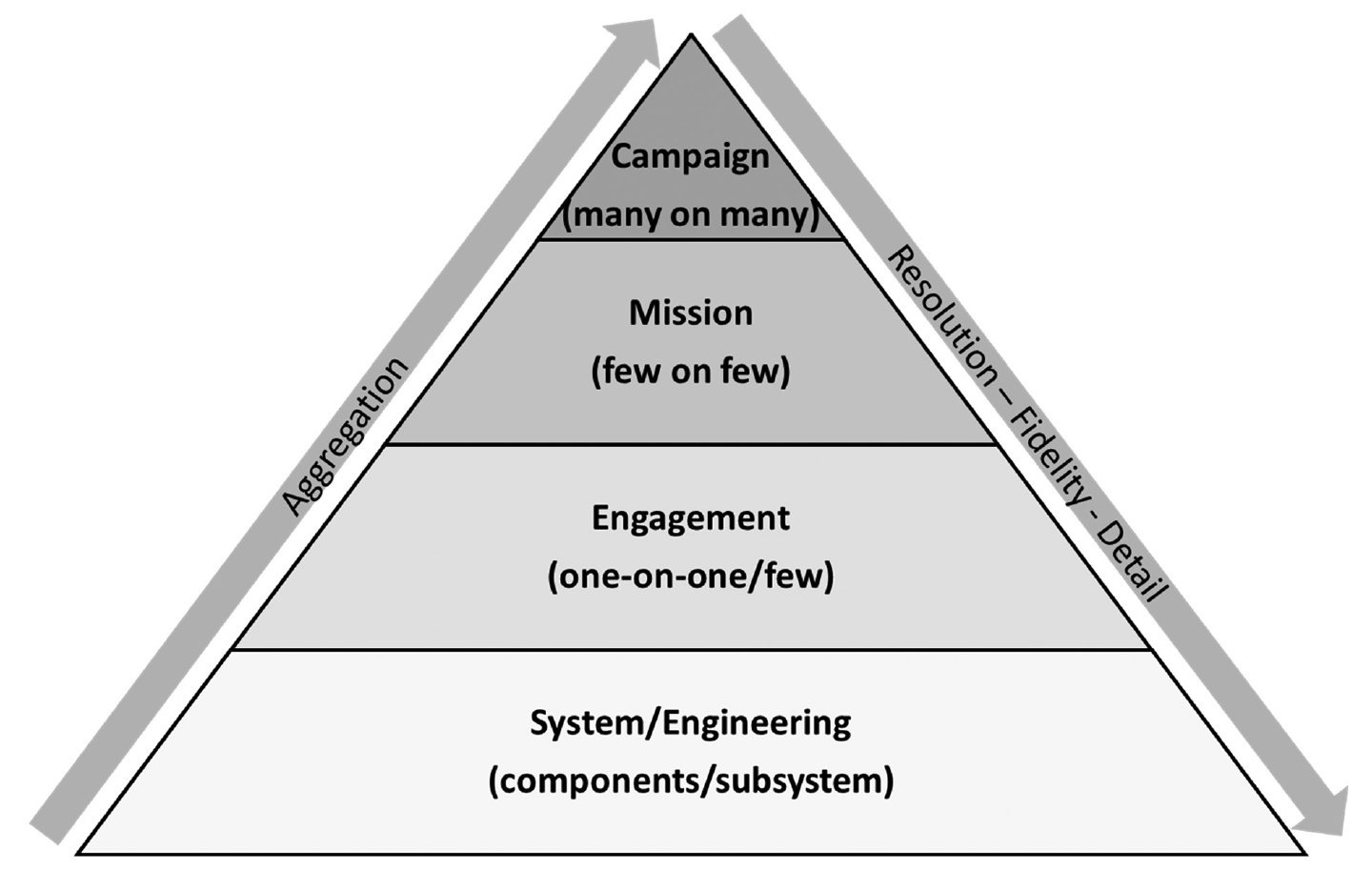}
\caption{Department of Defense Modeling and Simulation Hierarchy \citep{wade20}.}
\label{fig:hierarchy}
\end{figure}

Campaign analysis sits atop the military modeling hierarchy (Figure \ref{fig:hierarchy}). The campaign level trades off some detail to yield aggregated insights for military senior decision makers. Combat simulations and wargaming are two tools used individually or in tandem to study campaign analysis problems \citep{turn22}. Wargames are ideal to explore human decision making while simulations are ideal to deliver quantitative assessments and results under a typically  fixed operational plan \citep{turn22}. These tools are vital, yet take substantial time. The campaign wargame is personnel intensive, often requiring weeks (including rehearsals and participant travel) to develop one operational plan for one campaign setting \citep{burns15}. The combat simulation takes months to instantiate and assesses one plan at a time \citep{sweet20}. As a result, these methods cannot be scaled to develop a plan for many scenarios while accounting for the behavioral and uncertain dynamics of military campaigns. To address these limitations, we propose a stochastic game model to quickly develop initial operational plans and evaluate campaign inputs. 

While the stochastic game framework has been applied in the overall military domain, we did not identify any direct application to the campaign analysis level in the open literature. We assess that existing efforts are largely for the system/engineering or engagement levels with a few mission-level applications. At the system/engineering and engagement levels, \cite{ho22} review many of stochastic game applications. For instance, \cite{bach11} and \cite{he21} study the competitive interaction between radars and jammers. \cite{deli17} study power allocation for a radar network against multiple jammers. Then, \cite{bogd18} and \cite{gu11} study target selection and tracking in radar networks. Finally, \cite{krish08} study transmission in ground sensor networks. Stochastic games are also used for mission-level problems, involving military battles between forces. \cite{chang22} employ a partially observed stochastic game to assess modern combat in Mosul. \cite{mce04} apply a stochastic game to air operation command and control. These models consider a fixed role for each player (e.g., attacker versus defender, radar versus jammer, air force versus ground targets), while the role of opposing forces in a military campaign is often context-dependent. For instance during the World War II European campaign, Axis forces were simultaneously on the offensive or defensive. Furthermore, these models do not consider the operational level's interconnected nature of commands and domains.

Other game theoretic models have been studied for settings where opposing forces seek to gain military objectives by assigning resources to a subset of objectives. \cite{bier07} develop a sequential game to model a defender who first allocates defensive resources against a subsequent attack from an unknown attacker. In the Colonel Blotto game framework \citep{shub81,cough92,wash14}, zero-sum (static) matrix games are introduced where opposing military commanders simultaneously assign a fixed number of forces (e.g., individual soldiers) across a set of military objectives. The commander who places the highest number of forces on an objective gains control of that objective. Many extensions have been investigated, including asymmetry of resources \citep{chowd13} and objective heterogeneity  \citep{kove21}. \cite{Kaminsky84} and \cite{khar22} have extended the battle outcomes to be stochastic with a winning probability proportional to the amount of assigned forces. Such games typically assume players can assign resources to any objective, and cannot be applied to settings involving constraints rising from logistics lines or operational-level command structure. Our model also differs in that we consider the inherent differences between a player maintaining control of an objective they control and gaining control of an objective the player's adversary controls. Finally, we account for the interdependent nature of warfare, where controlling one objective changes the future transition dynamics for other objectives.

\cite{haywood54}, who reviews the World War II Battles of the Bismarck Sea (U.S. versus Japan) and Avranches-Gap (Allies versus Germany), provides a launching point for our model. He, and later \cite{cant03} and \cite{fox16}, demonstrate that portraying military battle  decision-making as a matrix game is a powerful decision support tool. While \cite{haywood54} provides a clear delineation of military decision making for the single-stage single-battle problem, we aim to extend his work by building a military campaign that accounts for the dynamic sequence of simultaneous stochastic battles conducted across multiple domains, linked through an interdependent transition model.

In summary, current military stochastic game applications often consider engineering or engagement level problems. On the other hand, existing game models that allocate resources to military objectives do not consider military logistics precedence, command structure, or the interdependence and dynamics of sequential battles. We contend that our work is the first application of the stochastic game framework to the study of dynamic, multi-battle, military campaigns.

%!TEX root = NRL Manuscript.tex

\section{Problem Description}\label{sec:Pb}

In this section, we define the campaign between two military forces. Then, we model their dynamic and stochastic interactions using a two-person, discounted, zero-sum stochastic game.

\subsection{Campaign Model}
We consider a competitive operational campaign between two players that represent opposing military forces. The campaign consists of a set $\O$ of common \emph{objectives} that the players aim to control. Each objective is defined as ``a decisive and attainable goal toward which every operation is directed" \citepalias{jp1-02}, and may represent controlling operational-level terrain (e.g., capital city, airport, canal) or a non-geographic goal such as gaining air superiority. To account for precedence constraints between military goals during a campaign, we partition objectives into a collection  $\X$ of \emph{axes}. Specifically, each axis $x \in \X$ is a totally ordered set of objectives, represented as $x = (o^x_1,\dots,o^x_{|x|})$, where $o^x_1$ (resp. $o^x_{|x|}$) represents the \emph{front} (resp. \emph{rear}) of axis $x$. As an example of a precedence relationship, consider that a military may not gain air superiority until they have neutralized air defenses and achieved air parity first.

Each player $i \in \{1,2\}$ manages a collection $\C^i$ of commanders, where each commander $c \in \C^i$ is \emph{responsible} for a subset $\X_c\subseteq \X$ of axes. We denote as $\O_c \coloneqq \{o \in x, \ \forall x \in \X_c\} \subseteq \O$ the objectives under the responsibility of commander $c$. We assume that each axis $x \in \X$ is under the responsibility of one commander $c \in \C^i$ from each player $i$, that is, $x \in \X_{c}$. We further assume that responsibilities are \emph{symmetric}, that is, for every $c^1 \in \C^1$, there exists $c^2 \in \C^2$ such that $\X_{c^1} = \X_{c^2}$.
%and that responsibilities are \emph{symmetric}, that is, commanders for each player vie for control of the same objectives. Mathematically, this translates as assuming that for $x \in \X$, there exist $c^1 \in \C^1$ and $c^2 \in \C^2$ such that $x \in \X_{c^1}$ and $x\in \X_{c^2}$.
%
%
%each who are \emph{responsible} for subsets of axes, with the assumption that each axis is under the responsibility of one commander from each player. We further assume that responsibilities are \emph{symmetric}, that is, commanders for each player vie for control of the same objectives. 
%Furthermore, for each commander $c \in \C$, we denote as $\X_c\subseteq \X$ the subset of axes under the responsibility of commander $c$ for each player, and as $\O_c = \{o \in x, \ \forall x \in \X_c\}$ the associated subset of objectives. 
The collections of commanders and their associated responsibilities represent the operational command structure.

Each player $i \in \{1,2\}$ operates a collection of bases $\mathcal{B}^i$, with each base managed by one of Player $i$'s commanders. For Player 1 (resp. Player 2), we assume that each commander's base provides access to the first objective $o^x_1$ (resp. last objective $o^x_{|x|}$) of each axis $x$ under their responsibility. A base, which may represent one geographic location (e.g. a port), multiple geographic locations (e.g., a dispersed Army Corps), or a non-geographic location (e.g., in the Cyber domain), represents the origin for operational command and control (C2), unit arrival, and the supply chain for the associated commander.

\subsection{Stochastic Game Formulation}

To capture the competitive and dynamic interactions of armed conflict, we formulate a two-player discounted zero-sum stochastic game over an infinite horizon, expressed by the tuple $\Gamma \coloneqq \langle\S,(\A^1,\A^2),P,L,\gamma\rangle$. The game is played over stages, that is, discrete times where players take actions. $\S$ represents the finite state space, $\A^1$ (resp. $\A^2$) the state-dependent action space of Player 1 (resp. Player 2), $P$ the state transition probability function, $L$ the single-stage loss function, and $\gamma$ the discount factor between subsequent stages. We next detail each of these quantities. Following the game-theoretic convention, when considering Player $i$ (for $i \in \{1,2\}$), we refer to their opponent as Player $-i$.

\textbf{State space:}
At any given stage, we model the campaign state using a vector $s \in \{1,2\}^\O$, which represents the control of each objective. Specifically, for any objective $o \in \O$, $s_o = 1$ (resp. $s_o = 2$) if objective $o$ is controlled by Player 1 (resp. Player 2). We denote as $\S \coloneqq \{1,2\}^\O$ the campaign state space, equipped with a partial order $\preceq$, representing the component-wise inequality between states.

\textbf{Action spaces:}
At every stage, the players select actions to vie for control of objectives. This can be achieved by Player $i$ attempting to maintain control of their objectives or gain control of their opponent's objectives. Each player's action is comprised of \emph{orders} for each commander. We assume that each commander for each player can give one order to \textit{attack} (\emph{atk}) or \textit{reinforce} (\emph{rfc}) a single objective under their responsibility. The order for all other objectives is \textit{none}. An attack order on objective $o \in \O$ gives the attacking player a chance of gaining control of $o$. A reinforce order on objective $o$  adds a defensive layer and reduces the chance of losing an objective should the opponent attack $o$. 

% Recall that these bases are not necessarily geographic entities; in the air example, a base projects air power, but is not constrained to any one airport. 

Let us consider a campaign state $s \in \S$. To attack or reinforce an objective, we assume the military operation requires C2, units, assets, and supplies such as fuel and ammunition that project from a base along the objective's axis. A military operation cannot bypass any interim objective on the corresponding axis, because these objectives are necessary to maintain the C2, units, assets, and supplies required to proceed to a subsequent objective. The military concept of \textit{line of communication} (LoC) historically originates from human-delivered communication, but today the concept extends beyond communication. While maintaining the acronym, we will use the terminology \textit{line of control}, which is a route that connects a military unit with its supply chain. A LoC is deemed ``open" if the military controls every objective along that route. In our setting, this translates as follows: Given an objective $o^x_{k}$ belonging to an axis $x \in \X$, Player 1 (resp. Player 2) has an open LoC to $o^x_{k}$ if $s_{o^x_j} = 1$ for every $j \in \{1,\dots,k-1\}$ (resp. $s_{o^x_j} = 2$ for every $j \in \{k+1,\dots,|x|\}$).  We denote as $\overline{\O}^i_s \subseteq \O$ the set of objectives with an open LoC from Player $i$'s bases $\mathcal{B}^i$. As a result, Player $i$ can reinforce an objective $o \in \O$ they control, provided it has an open LoC from $\mathcal{B}^i$ (i.e., if $o \in  \overline{\O}^i_s$ and $s_o = i$). Furthermore, Player $i$ can attack an objective they do not control provided it has an open LoC from $\mathcal{B}^i$  (i.e., if $o \in  \overline{\O}^i_s$ and $s_o = -i$).

For each $i \in \{1,2\}$, we denote Player $i$'s action as $a^i \in \{\textit{atk,rfc,none}\}^{\O}$, where $a^i_o$ represents the order for objective $o \in \O$. Action $a^i$ is feasible if it satisfies
\begin{align}
&\forall o \in \O, \ a_o^i \in \begin{cases} \{\textit{none}\} & \text{if } o \notin \overline{\O}^i_s\\ \{\textit{none,rfc}\}& \text{if } o \in \overline{\O}^i_s \text{ and } s_o = i\\  \{\textit{\textit{none,atk}}\}& \text{if } o \in \overline{\O}^i_s \text{ and } s_o = -i\end{cases}\label{actions_1}\\
&\forall c \in \C^i, \ |\{a^i_o \in \{\textit{atk,rfc}\}, \ \forall o \in \O_c\}| \leq 1.\label{actions_2}
\end{align}
We denote as $\A^i_s$ the set of feasible actions for Player $i$ at state $s$.

\begin{figure}[htbp]
        \centering
        \small
        \begin{tikzpicture}[x = 1.2cm,y=1.3cm]
%        \node[text=black](d1) at (-0.7,2.3) {$0.6$};

	\node[shape=rectangle,draw=black,fill=blue,text=black,inner sep = 0.25cm] (b11) at (-1,3) {};
	\node[shape=circle,draw=black,fill=red,text=black,inner sep = 0.1cm] (1) at (0,3) {$1$};
	\node[shape=circle,draw=black,fill=red,text=black,inner sep = 0.1cm] (2) at (1,3) {$2$};
	\node[shape=rectangle,draw=black,fill=red,text=black,inner sep = 0.25cm] (b21) at (2,3) {};
	
	\node[shape=rectangle,draw=black,fill=blue,text=black,inner sep = 0.25cm] (b12) at (-1,1.5) {};
	\node[shape=circle,draw=black,fill=blue,text=white,inner sep = 0.1cm] (3) at (0,2) {$3$};
	\node[shape=circle,draw=black,fill=red,text=black,inner sep = 0.1cm] (4) at (1,2) {$4$};
	\node[shape=circle,draw=black,fill=blue,text=white,inner sep = 0.1cm] (5) at (0,1) {$5$};
	\node[shape=circle,draw=black,fill=blue,text=white,inner sep = 0.1cm,double,thick] (6) at (1,1) {$6$};
	\node[shape=rectangle,draw=black,fill=red,text=black,inner sep = 0.25cm] (b22) at (2,1.5) {};
	
	\path (b11) edge  (1);
	\path (1) edge  (2);
	\path (2) edge  (b21);
	\path (b12) edge  (3);
	\path (3) edge  (4);
	\path (4) edge  (b22);
	\path (b12) edge  (5);
	\path (5) edge  (6);
	\path (6) edge  (b22);
	
	\node at (-4,2.25) {\color{blue} Player 1};
	\node at (5,2.25) {\color{red} Player 2};
	
	\node (b1)at (-1.4,2.25) {\color{blue} $\mathcal{B}^1$};
	\node (b2) at (2.4,2.25) {\color{red} $\mathcal{B}^2$};
	
	\path[->,>=stealth',shorten >=1pt,thick,blue] (b1) edge  (b11);
	\path[->,>=stealth',shorten >=1pt,thick,blue] (b1) edge  (b12);
	
	\path[->,>=stealth',shorten >=1pt,thick,red] (b2) edge  (b21);
	\path[->,>=stealth',shorten >=1pt,thick,red] (b2) edge  (b22);

	\node at (-2.5,3) {\color{blue} Commander 1};
	\node at (3.5,3) {\color{red} Commander 1};
	\node at (-2.5,1.5) {\color{blue} Commander 2};
	\node at (3.5,1.5) {\color{red} Commander 2};
	\draw (0,3) node[cross=7.5pt, blue, line width = 0.5mm,scale=0.15] {};    
		
    \end{tikzpicture}
\caption{Campaign example with 6 objectives and 2 bases for each player. Player 1 controls objectives 3, 5, and 6. Player 2 controls objectives 1, 2, and 4. Player 1 orders Commander 1 to attack objective 1 and Commander 2 to reinforce objective 6.}
\label{fig:example}
    \end{figure}
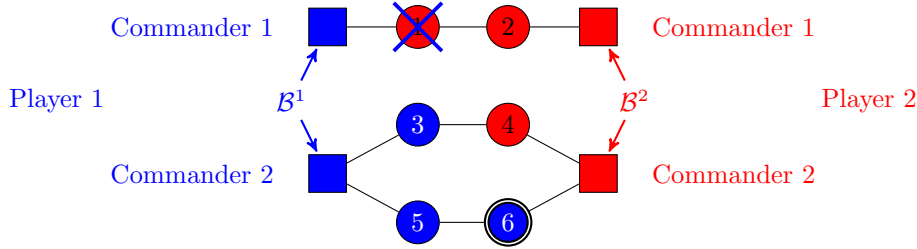

In Figure \ref{fig:example}, we illustrate a campaign containing 6 objectives partitioned into three axes $x_1 = \{1,2\}$, $x_2 = \{3,4\}$, $x_3 = \{5,6\}$ managed by 2 commanders for each player. For both players, Commander 1 is responsible for $x_1$ and Commander 2 is responsible for $x_2$ and $x_3$. The current state is given by  $s = (2,2,1,2,1,1)$. In this example, Player 1 selects a feasible action $a^1 = (\textit{atk},\textit{none},\textit{none},\textit{none},\textit{none},\textit{rfc})$ that attacks objective 1 and reinforces objective 6.

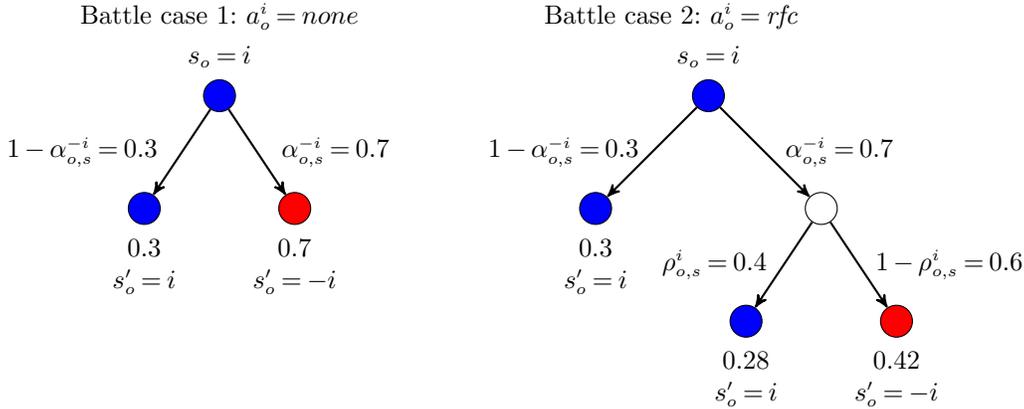
\begin{figure}[htbp]
        \centering
        \small
        \begin{tikzpicture}[->,>=stealth',shorten >=0pt,x = 1.0cm,y=1.5cm]
        
	\node[shape=circle,draw=black,fill=blue,text=black,inner sep = 0.15cm] (1) at (0,0) {};
	\node[shape=circle,draw=black,fill=blue,text=black,inner sep = 0.15cm] (2) at (-1,-1) {};
	\node[shape=circle,draw=black,fill=red,text=black,inner sep = 0.15cm] (3) at (1,-1) {};

	\path[thick] (1) edge (2);
	\path[thick] (1) edge (3);
	
	\node at (0,0.35) {$s_o = i$};
	\node[anchor=west] at (0.7,-0.5) {$\alpha^{-i}_{o,s} = 0.7$};
	\node[anchor=east] at (-0.7,-0.5) {$1-\alpha^{-i}_{o,s} = 0.3$};
	
	\node at (-1,-1.35) {$0.3$};
	\node at (1,-1.35) {$0.7$};
	
	\node at (-1,-1.65) {$s_o^\prime = i$};
	\node at (1,-1.65) {$s_o^\prime = -i$};
	 
	 \node at (0,0.7) {Battle case 1: $a_o^{i} = none$};

	 \node at (6,0.7) {Battle case 2: $a_o^{i} = \textit{rfc}$};

	\node[shape=circle,draw=black,fill=blue,text=black,inner sep = 0.15cm] (4) at (6.5,0) {};
	\node[shape=circle,draw=black,fill=blue,text=black,inner sep = 0.15cm] (5) at (5,-1) {};
	\node[shape=circle,draw=black,text=black,inner sep = 0.15cm] (6) at (8,-1) {};

	\path[thick] (4) edge (5);
	\path[thick] (4) edge (6);
	
	\node at (6.5,0.35) {$s_o = i$};
	\node[anchor=west] at (7.4,-0.5) {$\alpha^{-i}_{o,s} = 0.7$};
	\node[anchor=east] at (5.7,-0.5) {$1-\alpha^{-i}_{o,s} = 0.3$};
	
	\node at (5,-1.35) {$0.3$};
%	\node at (8,-1.35) {$0.7$};
	
	\node at (5,-1.65) {$s_o^\prime = i$};
%	\node at (8,-1.65) {$s_o^\prime = -i$};

	 \node[shape=circle,draw=black,fill=blue,text=black,inner sep = 0.15cm] (7) at (7,-2) {};
	\node[shape=circle,draw=black,fill=red,text=black,inner sep = 0.15cm] (8) at (9,-2) {};
	 
\path[thick] (6) edge (7);
	\path[thick] (6) edge (8);
		
		\node[anchor=west] at (8.6,-1.5) {$1-\rho^{i}_{o,s} = 0.6$};
	\node[anchor=east] at (7.4,-1.5) {$\rho^{i}_{o,s} = 0.4$};

		\node at (7,-2.35) {$0.28$};
	\node at (9,-2.35) {$0.42$};
	
	\node at (7,-2.65) {$s_o^\prime = i$};
	\node at (9,-2.65) {$s_o^\prime = -i$};

    \end{tikzpicture}
\caption{Probabilities of battle outcomes when Player $-i$ attacks objective $o$.}
\label{fig:gates}
    \end{figure}

\textbf{Transition probabilities:}
If an objective is attacked by the non-controlling player, a \textit{battle} occurs with an uncertain outcome. We define a transition probability function $P$ that, given a current state and the players' actions, determines the probability of transitioning to a new state. These probabilities are created by domain experts including wargamers, operational planners, and weapon system experts. In this article, we model the dynamics of a battle for an objective as a probability chain with one or two events, where the second event only occurs if the objective is reinforced by the controlling player. Specifically, consider a battle occurring for an objective $o \in \O$ controlled by Player $i$ (i.e., $s_o = i$, $a^{-i}_o = atk$). Player $i$ is the \emph{defender} of $o$, and Player $-i$ is the \emph{attacker}. There are two possible outcomes for the subsequent state $s^\prime_o$ of that objective: the objective is retained by the defender (i.e., $s^\prime_o = i$), or the objective is gained by the attacker (i.e., $s^\prime_o = -i$). If $a^i_o = none$ (i.e., the defender does not reinforce objective $o$), then Player $i$ still defends the objective $o$ with a first level of defense (e.g., with units already assigned to the objective), and Player $-i$ gains control of objective $o$ with a state-dependent \emph{base attack success probability} $\alpha^{-i}_{o,s} \in [0,1]$. If $a^i_o = \textit{rfc}$ (i.e., the defender does reinforce objective $o$), then Player $i$ adds another defensive layer, which itself manages to thwart the attack with a state-dependent \emph{reinforce success probability} $\rho^i_{o,s} \in [0,1]$. A reinforce order indicates that a commander dedicates \textit{additional} units and assets along the LoC to protect the objective. Assuming independence between the two defensive layers, Player $-i$'s attack becomes successful with probability $\alpha^{-i}_{o,s} \cdot (1-\rho^i_{o,s})$. Figure \ref{fig:gates} illustrates the transition dynamics for one battle. 

For notational convenience, we let $\alpha^i_{o,s} = 1$ if $s_o = i$ and $\rho^i_{o,s} = 0$ if $s_o = -i$. Through $\alpha$ and $\rho$, controlling objectives in one axis may affect operations across all axes. For instance gaining air superiority, although perhaps not a precedence requirement before launching a ground attack, will greatly increase its likelihood of succeeding.

We make the following natural assumptions on the base attack and reinforce success probabilities:
\begin{assumption} \label{ass:atk}

Player 1's (resp. Player 2's) base attack and reinforce success probabilities are antitone (resp. isotone) functions of the state space:
\begin{align*}
&\forall s \preceq s^\prime \in \S, \ \forall o \in \O, \ \alpha^1_{o,s} \geq \alpha^1_{o,s^\prime}, \ \rho^1_{o,s} \geq \rho^1_{o,s^\prime}, \text{ and } \  \alpha^2_{o,s} \leq \alpha^2_{o,s^\prime}, \ \rho^2_{o,s} \leq \rho^2_{o,s^\prime}.
\end{align*}

\end{assumption}

\begin{assumption} \label{ass:rfc}
For every player, the probability of gaining control of a reinforced objective is at most the probability of maintaining control of that reinforced objective, given an identical control of all other objectives:
\begin{align*}
&\forall i \in \{1,2\}, \ \forall o \in \O, \ \forall (s,s^\prime) \in \S^2 \ | \ s_{o^\prime}=s'_{o^\prime} \ \forall o^\prime \in \O\setminus\{o\}, \text{ then } \ 1-\alpha^{-i}_{o,s}(1-\rho^i_{o,s}) \geq \alpha^i_{o,s^\prime}(1-\rho^{-i}_{o,s^\prime}).
\end{align*}

\end{assumption}

Assumption \ref{ass:atk} follows the military intuition that one force can better attack or reinforce if they control more objectives. For instance, an air mission will have higher likelihood of success if multiple airbases are controlled. Similarly, a ground operation will be more successful if multiple key terrain objectives are already controlled. We do not model the notion of spreading forces too thin and assume that the players' commanders have sufficient combat power in their bases such that it is never a disadvantage to control more objectives.

Assumption \ref{ass:rfc} is rooted in military doctrine that more forces are required to attack than to defend an objective. In ancient times, Sun Tzu \citep{suntzu13} advocated a five to one ratio for attacking forces versus defending forces. More recently the U.S. Army \citepalias{fm6-0} recommends a three to one force ratio for an attacker versus a prepared defense. In the case study (Section \ref{sec:soliciting_T}), we discuss one process to create $\alpha$ and $\rho$.

When multiple battles occur simultaneously, we assume that their outcomes are independent. Then, the transition probability function is given as follows:
\begin{align*}
&\forall s \in \S, \ \forall (a^1,a^2) \in \A^1_s \times \A^2_s, \ \forall s^\prime \in \S, \ P(s^\prime \, | \, s,a^1,a^2) = \prod_{o \in \O}p_o(s_o^\prime \, | \, s, a^1_o,a^2_o),
\end{align*}
where,
\begin{align*}
\forall i \in \{1,2\}, \ \forall o \in \O \ | \ s_o^\prime = i, \ p_o(s_o^\prime \, | \, s, a^1_o,a^2_o)  = \begin{cases}
1 - \alpha^{-i}_{o,s}\cdot\mathds{1}_{\{a^{-i}_o = \textit{atk}\}} \cdot(1-\rho^i_{o,s}\cdot\mathds{1}_{\{a^i_{o} = \textit{rfc}\}})       &\text{if } s_o=\phantom{-}i\\
\alpha^{i}_{o,s}\cdot\mathds{1}_{\{a^{i}_o = \textit{atk}\}} \cdot(1-\rho^{-i}_{o,s}\cdot\mathds{1}_{\{a^{-i}_{o} = \textit{rfc}\}})       &\text{if } s_o=-i.
\end{cases}
\end{align*}

\textbf{Stage loss:}
We suppose that each objective $o \in \O$ is associated with an individual loss value $\ell_o \in \mathbb{R}_{\geq0}$, which is incurred by Player 1 when objective $o$ is controlled by Player 2. Then, the total stage loss incurred at state $s \in \S$ is given by  $L(s) = \sum_{o \in \O} \ell_o\cdot \mathds{1}_{\{s_o=2\}}$. Player 1 seeks to minimize the loss, while Player 2 seeks to maximize it.

\textbf{Discount factor:}
We suppose that the players interact dynamically over an infinite horizon, and we consider a discount factor $\gamma \in (0,1)$ that impacts the importance of future stage losses relative to the current loss \citep{lesl20}. Another interpretation is that $1 - \gamma$ represents the probability that the conflict ends at the end of each stage \citep{pute08}. For our application, $\gamma$ models the incentive for both sides to conclude a campaign expediently, before casualties and equipment losses mount. Since modeling casualties and equipment losses explicitly would necessitate a prohibitively large state space, we offer that a smaller discount factor accounts for attrition implicitly, as gaining objectives at a later time will have less utility due to attrition from a lengthy campaign. From a geopolitical perspective, a smaller discount factor may represent the importance of concluding a conflict rapidly before political will subsides. We assume that concluding a conflict is equally preferable for both sides.

\textbf{Stationary policies:}
At any state, each player may benefit from randomizing their actions to create uncertainty that cannot be exploited by their opponent. We focus our analysis on \emph{stationary} (or \emph{Markovian}) \emph{policies} for each player. A stationary policy $\pi^i$ for Player $i$ maps each state $s \in \S$ to a probability distribution $\pi^i(s)$ over Player $i$'s state-dependent action set $\A^i_s$. We denote as $\Delta(\A^i_s)$ the probability simplex over $\A^i_s$ and $\Delta^i \coloneqq \prod_{s \in \S} \Delta(\A^i_s)$ the set of stationary  policies for Player $i$. Given a policy $\pi^i \in \Delta^i$ for  Player $i$, $\pi^i(s) \in \Delta(\A^i_s)$ denotes the mixed strategy implemented at a given state $s$ and $\pi^i(s,a^i)$ the probability that action $a^i \in \A^i_s$ is realized at state $s$. Furthermore, for every state $s \in \S$, we assume that the probability distributions $\pi^1(s)$ and  $\pi^2(s)$ are independent.

\textbf{Cumulative loss:}
Given a policy profile $(\pi^1,\pi^2) \in \Delta^1\times \Delta^2$, the stochastic game $\Gamma$ starts from an initial state $s^{(0)} = s\in \S$. Then, each player $i$ observes $s^{(0)}$ and simultaneously draws an action $a^{i,(0)} \sim \pi^i(s^{(0)})$. The campaign then transitions to a new state $s^{(1)}\sim P(\cdot \, | \, s^{(0)},a^{1,(0)},a^{2,(0)})$. The process is infinitely repeated, resulting in a Markov chain $s^{(0)}, s^{(1)},\dots$ Given the discount factor $\gamma$, the corresponding expected discounted cumulative loss is then given by:
\begin{align*}
V(s,\pi^1,\pi^2) = \mathbb{E}\left[\sum_{t=0}^{+\infty} \gamma^t \cdot L(s^{(t)}) \, \Big | \, s^{(0)} = s\right  ],
\end{align*}
where the expectation is taken over the players' actions $a^{i,(t)} \sim \pi^i(s^{(t)})$ and the state transitions $s^{(t+1)} \sim P(\cdot \, | \, s^{(t)},a^{1,(t)},a^{2,(t)})$ at each stage $t$. We refer to $V(s,\pi^1,\pi^2)$ as the \emph{value} function under strategy profile $(\pi^1,\pi^2)$ and initial state $s$. Equivalently, the value function is given by the following recursive Bellman policy equation:
\begin{align*}
\ V(s,\pi^1,\pi^2) = \sum_{\substack{
a^1\in\A^1_s \\
a^2\in\A^2_s
}} \pi^1(s,a^1)\cdot \pi^2(s,a^2) \cdot\left(L(s) + \gamma \cdot \sum_{s^\prime \in \S} P(s^\prime \, | \, s, a^{1},a^{2}) \cdot V(s^\prime,\pi^1,\pi^2)\right).
\end{align*}
%\begin{align*}
%\forall s \in \S, \ V(s,\pi^1,\pi^2) = L(s) + \gamma \sum_{a^1\sim\pi^1(s)}\sum_{a^2\sim\pi^2(s)} \pi^1(s,a^1)\cdot \pi^2(s,a^2) \sum_{s^\prime \in \S} P(s^\prime \ | \ s, a^{1},a^{2}) \cdot V(s^\prime,\pi^1,\pi^2).
%\end{align*}
Given a starting state $s$, Player $1$ (resp. Player $2$) aims to minimize (resp. maximize) the value function.

\textbf{Markov perfect equilibria:}
In such stochastic games, the optimal solution concept is given by \emph{Markov perfect equilibria} (MPE). Specifically, a policy profile $(\pi^{1^*},\pi^{2^*}) \in \Delta^1\times \Delta^2$ is an MPE of the stochastic game $\Gamma$ if
\begin{align*}
\forall s \in \S, \ \forall (\pi^{1},\pi^{2}) \in \Delta^1\times \Delta^2, \ V(s,\pi^{1^*},\pi^{2}) \leq V(s,\pi^{1^*},\pi^{2^*}) \leq  V(s,\pi^{1},\pi^{2^*}).
\end{align*}
In other words, no player has a unilateral incentive to deviate from their policy regardless of the initial state. For every $s \in \S$, we denote as $V^*(s) \coloneqq V(s,\pi^{1^*},\pi^{2^*})$ the optimal value of the game $\Gamma$ at state $s$.  By extension, the approximate solution concept is given by approximate MPE: Given $\epsilon \in \mathbb{R}_{>0}$, a policy profile $(\pi^{1^\prime},\pi^{2^\prime}) \in \Delta^1\times \Delta^2$ is an $\epsilon$-MPE of the stochastic game $\Gamma$ if
\begin{align*}
\forall s \in \S, \ \forall (\pi^{1},\pi^{2}) \in \Delta^1\times \Delta^2, \ V(s,\pi^{1^\prime},\pi^{2}) -\epsilon \leq V(s,\pi^{1^\prime},\pi^{2^\prime}) \leq  V(s,\pi^{1},\pi^{2^\prime}) + \epsilon.
\end{align*}

In his fundamental paper, \cite{shap53} generalized \citeauthor{vNeu28}'s minimax theorem to show the existence of MPE in such discounted zero-sum stochastic games. The proof contains \emph{Shapley's value iteration}, which is an algorithm to approximately compute the optimal value function and optimal policies. Since Shapley, numerous algorithms have been proposed to find MPE or $\epsilon$-MPE solutions to discounted, zero-sum stochastic games \citep{rag91}. These methods include the Hoffman-Karp, Pollatschek-Avi Itzhak, van der Wal, and Brown's fictitious play algorithms \citep{hoff66,poll69,van78,vrieze82}. However, stochastic games face the well-known curse of dimensionality, which hinders the scalability of classical solution algorithms. In the next section, we exploit the campaign's structure to mitigate these computational issues and obtain MPEs more efficiently.

%!TEX root = NRL Manuscript.tex

\section{Solution Methodology} \label{sec:approaches}

In this section, we derive equilibrium properties of the stochastic game $\Gamma$. This permits us to significantly reduce the state and action spaces, and accelerate the value iteration algorithm for computing MPEs.

\subsection{Shapley's Value Iteration}\label{sec:VI}

 Shapley's classical value iteration (VI) algorithm is based on the fact that the optimal value function $V^* = (V^*(s))_{s \in \S}$ of the game $\Gamma$ is the fixed point of the contraction mapping $T: \mathbb{R}^\S \to \mathbb{R}^\S$ defined by $T(V^\prime) = (T(V^\prime,s))_{s \in \S}$ for every $V^\prime \in  \mathbb{R}^\S$ with 
\begin{align*}
\ T(V^\prime,s) \coloneqq \min_{\pi^1(s) \in \Delta(\A^1_s)} \max_{a^2 \in \A^2_s} \sum_{a^1\in \A^1_s} \pi^1(s,a^1) \cdot \left(L(s) + \gamma  \cdot \sum_{s^\prime \in \S} P(s^\prime \, | \, s, a^{1},a^{2}) \cdot V^\prime(s^\prime) \right).
\end{align*}

The contraction mapping $T(V^\prime)$ consists of solving at every state $s \in \S$ a zero-sum matrix game $\Gamma(V^\prime,s)$, where the payoff matrix is given by $R(V^\prime,s) = (R(V^\prime,s,a^1,a^2))_{(a^1,a^2) \in \A^1_s \times \A^2_s}$ with
\begin{align*}
\forall (a^1,a^2) \in \A^1_s \times \A^2_s, \ R(V^\prime,s,a^1,a^2) \coloneqq L(s) + \gamma \cdot  \sum_{s^\prime \in \S} P(s^\prime \, | \, s, a^{1},a^{2}) \cdot V^\prime(s^\prime).
\end{align*}
As a reminder, every zero-sum game $\Gamma(V^\prime,s)$ can be solved using the following linear program: 
\begin{alignat}{3}
\text{LP}(R(V^\prime,s),\A^1_s,\A^2_s): \quad \min_{\pi^1(s), \, z} \ & z \nonumber\\
\text{s.t.} \ & \sum_{a^1 \in \A^1_s}  R(V^\prime,s,a^1,a^2)\cdot \pi^1(s,a^1)\leq z, \ &&\forall a^2 \in \A^2_s\nonumber\\
& \sum_{a^1 \in \A^1_s} \pi^1(s,a^1)  = 1\nonumber\\
& \pi^1(s,a^1) \geq 0, \ &&\forall a^1 \in \A^1_s.\nonumber
\end{alignat}

Its optimal primal and dual solutions provide Player $1$'s and Player $2$'s respective mixed strategies in equilibrium of $\Gamma(V^\prime,s)$, and its optimal value provides $T(V^\prime,s)$.

The VI algorithm first initializes with $V^{(0)} \in \mathbb{R}^\S$ (we select $V^{(0)} = L$) and iteratively applies the mapping $T$, thus creating a sequence of vectors $(V^{(t)})_{t \in \mathbb{Z}_{\geq0}}$ that converges towards $V^*$. Given $\epsilon \in \mathbb{R}_{>0}$, the algorithm terminates when $\|V^{(t)}- V^{(t-1)}\|_{\infty} \leq \epsilon (1-\gamma)/(2\gamma)$. This ensures that $\|V^{*}- V^{(t)}\|_{\infty} \leq \epsilon/2$ and that solving the zero-sum game $\Gamma(V^{(t-1)},s)$ for every state $s$ provides an $\epsilon$-MPE $(\pi^{1^\prime},\pi^{2^\prime}) \in \Delta^1\times \Delta^2$ \citep{pute08,10.1093/nsr/nwac256}.  The algorithm is guaranteed to terminate with $t \leq \lceil\frac{\log(\epsilon \cdot(1-\gamma)^2) - \log(2\sum_{o \in \O}\ell_o)}{\log \gamma}\rceil$. Since the running time of this algorithm is primarily driven by the number of states and actions, we next analyze the game's structure to reduce the state and action spaces and improve the algorithm's efficiency.

\subsection{State and Action Space Reduction} \label{sec:state_reduction}

We recall that the stochastic game $\Gamma$ is of large size, namely, its state space is of size $|\S| = 2^{|\O|}$, and Player $i$'s action space at every state $s \in \S$ is of size $|\A^i_s| = \prod_{c \in \C^i} (1 + |\overline{\O}^i_s \cap \O_c|)$. By leveraging the military operational C2 and logistics structure, we next derive equilibrium properties of $\Gamma$.

\begin{theorem}\label{thm:isotone}
    
The optimal value of $\Gamma$ is an isotone function of the state space and satisfies
%\begin{align}
%\forall s \prec s^\prime \in \S, \ V^*(s) < V^*(s^\prime).\label{new:value_isotone}
%\end{align}
\begin{align}
\forall s\preceq s^\prime \in \S, \ V^{*}(s^\prime) - V^{*}(s) \geq  \sum_{o \in \O} \ell_o \cdot \mathds{1}_{\{s^\prime_o = 2 \text{ and } s_o = 1\}}. \label{new:value_isotone}
\end{align}
\end{theorem}

From Theorem \ref{thm:isotone}, we observe that if the initial state is such that more objectives are controlled by Player 1, then the cumulative loss incurred over the infinite horizon in equilibrium will be lower. While this result seems intuitive, its proof is more involved, as typical arguments to show monotonicity in Markov decision processes do not hold \citep{pute08}. The challenge lies in that the state space only has a partial order, the action spaces are state-dependent, and the likelihood of transitioning to a better state is not an isotone function of the current state for every action. Instead, our proof exploits the LoC structure, the game-theoretic interactions between players, and Assumptions \ref{ass:atk} and \ref{ass:rfc} on the characteristics of the transition probability function. In fact, the following counterexample shows that the monotonicity property does not necessarily hold in equilibrium without Assumption \ref{ass:rfc}.

\begin{counterexample}

Consider a campaign with three objectives $\{1,2,3\}$ illustrated in Figure \ref{fig:counter_example}. Each objective is on its own axis and is managed by a distinct commander. The objective losses satisfy $\ell_1 > 0$  and $\ell_2 = \ell_3$, and all reinforce success probabilities at every state are 0.

We next consider two possible initial states given by $s = (1,1,2)$ and $s^\prime = (2,1,2)$, which satisfy $s \preceq s^\prime$. We assume that Player 1's base attack success probabilities satisfy $\alpha^1_{1,s^\prime} = 1$, $\alpha^1_{2,s}= \alpha^1_{2,s^\prime} = 0$, and $\alpha^1_{3,s} = \alpha^1_{3,s^\prime} =1$. Similarly, Player 2's base attack success probabilities satisfy $\alpha^2_{1,s} = 1$, $\alpha^2_{2,s}= \alpha^2_{2,s^\prime} = 1$, and $\alpha^2_{3,s} = \alpha^2_{3,s^\prime} =0$.

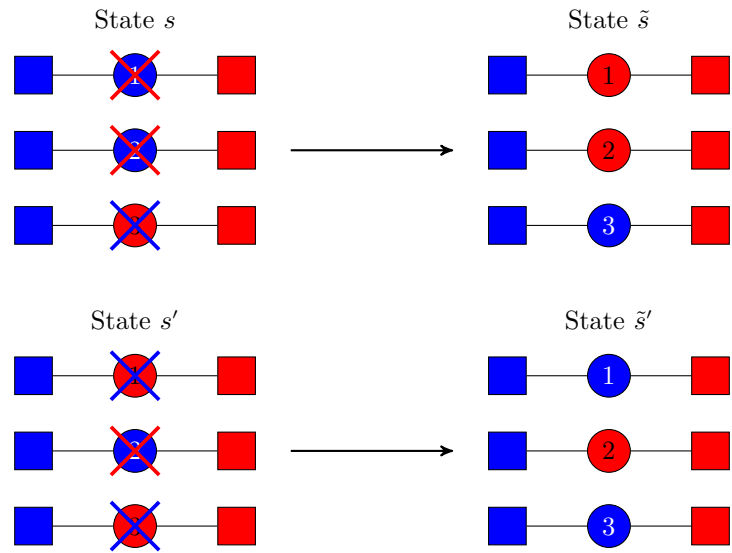
\begin{figure}[htbp]
        \centering
        \small
        \begin{tikzpicture}[x = 0.9cm,y=1.0cm]
%        \node[text=black](d1) at (-0.7,2.3) {$0.6$};

	\def\yspace{4}
	\def\xspace{7}
	
	\node at (0.5,3.75) {State $s$};

	\node[shape=rectangle,draw=black,fill=blue,text=black,inner sep = 0.25cm] (b11) at (-1,3) {};
	\node[shape=circle,draw=black,fill=blue,text=white,inner sep = 0.1cm] (1) at (0.5,3) {$1$};
%	\node[shape=circle,draw=black,fill=blue,text=white,inner sep = 0.1cm] (2) at (1,3) {$2$};
	\node[shape=rectangle,draw=black,fill=red,text=black,inner sep = 0.25cm] (b21) at (2,3) {};
	
	\node[shape=rectangle,draw=black,fill=blue,text=black,inner sep = 0.25cm] (b12) at (-1,2) {};
	\node[shape=circle,draw=black,fill=blue,text=white,inner sep = 0.1cm] (2) at (0.5,2) {$2$};
%	\node[shape=circle,draw=black,fill=red,text=black,inner sep = 0.1cm] (4) at (1,2) {$4$};
	\node[shape=rectangle,draw=black,fill=red,text=black,inner sep = 0.25cm] (b22) at (2,2) {};
	
	\node[shape=rectangle,draw=black,fill=blue,text=black,inner sep = 0.25cm] (b13) at (-1,1) {};
%	\node[shape=circle,draw=black,fill=blue,text=white,inner sep = 0.1cm] (5) at (0,1) {$5$};
	\node[shape=circle,draw=black,fill=red,text=black,inner sep = 0.1cm] (3) at (0.5,1) {$3$};
	\node[shape=rectangle,draw=black,fill=red,text=black,inner sep = 0.25cm] (b23) at (2,1) {};

	\path (b11) edge  (1);
%	\path (1) edge  (2);
	\path (1) edge  (b21);
	\path (b12) edge  (2);
%	\path (3) edge  (4);
	\path (2) edge  (b22);
	\path (b13) edge  (3);
%	\path (5) edge  (6);
	\path (3) edge  (b23);
	
	\draw (0.5,3) node[cross=7.5pt, red, line width = 0.5mm,scale=0.15] {};    
	\draw (0.5,2) node[cross=7.5pt, red, line width = 0.5mm,scale=0.15] {};  
	\draw (0.5,1) node[cross=7.5pt, blue, line width = 0.5mm,scale=0.15] {};

	%%------- Bottom Left
	
	\node at (0.5,3.75-\yspace) {State $s^\prime$};

	\node[shape=rectangle,draw=black,fill=blue,text=black,inner sep = 0.25cm] (b11prime) at (-1,3-\yspace) {};
%	\node[shape=circle,draw=black,fill=blue,text=white,inner sep = 0.1cm] (1prime) at (0,3-\yspace) {$1$};
	\node[shape=circle,draw=black,fill=red,text=black,inner sep = 0.1cm] (1prime) at (0.5,3-\yspace) {$1$};
	\node[shape=rectangle,draw=black,fill=red,text=black,inner sep = 0.25cm] (b21prime) at (2,3-\yspace) {};
	
	\node[shape=rectangle,draw=black,fill=blue,text=black,inner sep = 0.25cm] (b12prime) at (-1,2-\yspace) {};
	\node[shape=circle,draw=black,fill=blue,text=white,inner sep = 0.1cm] (2prime) at (0.5,2-\yspace) {$2$};
%	\node[shape=circle,draw=black,fill=red,text=black,inner sep = 0.1cm] (4prime) at (1,2-\yspace) {$4$};
	\node[shape=rectangle,draw=black,fill=red,text=black,inner sep = 0.25cm] (b22prime) at (2,2-\yspace) {};
	
	\node[shape=rectangle,draw=black,fill=blue,text=black,inner sep = 0.25cm] (b13prime) at (-1,1-\yspace) {};
%	\node[shape=circle,draw=black,fill=blue,text=white,inner sep = 0.1cm] (5prime) at (0,1-\yspace) {$5$};
	\node[shape=circle,draw=black,fill=red,text=black,inner sep = 0.1cm] (3prime) at (0.5,1-\yspace) {$3$};
	\node[shape=rectangle,draw=black,fill=red,text=black,inner sep = 0.25cm] (b23prime) at (2,1-\yspace) {};

	\path (b11prime) edge  (1prime);
%	\path (1prime) edge  (2prime);
	\path (1prime) edge  (b21prime);
	\path (b12prime) edge  (2prime);
%	\path (3prime) edge  (4prime);
	\path (2prime) edge  (b22prime);
	\path (b13prime) edge  (3prime);
%	\path (5prime) edge  (6prime);
	\path (3prime) edge  (b23prime);
	
	\draw (0.5,3-\yspace) node[cross=7.5pt, blue, line width = 0.5mm,scale=0.15] {};    
	\draw (0.5,2-\yspace) node[cross=7.5pt, red, line width = 0.5mm,scale=0.15] {};  
	\draw (0.5,1-\yspace) node[cross=7.5pt, blue, line width = 0.5mm,scale=0.15] {};

	%%------- Top Right

	\node at (0.5+\xspace,3.75) {State $\tilde{s}$};

	\node[shape=rectangle,draw=black,fill=blue,text=black,inner sep = 0.25cm] (b11tilde) at (-1+\xspace,3) {};
%	\node[shape=circle,draw=black,fill=blue,text=white,inner sep = 0.1cm] (1tilde) at (0+\xspace,3) {$1$};
	\node[shape=circle,draw=black,fill=red,text=black,inner sep = 0.1cm] (1tilde) at (0.5+\xspace,3) {$1$};
	\node[shape=rectangle,draw=black,fill=red,text=black,inner sep = 0.25cm] (b21tilde) at (2+\xspace,3) {};
	
	\node[shape=rectangle,draw=black,fill=blue,text=black,inner sep = 0.25cm] (b12tilde) at (-1+\xspace,2) {};
	\node[shape=circle,draw=black,fill=red,text=black,inner sep = 0.1cm] (2tilde) at (0.5+\xspace,2) {$2$};
%	\node[shape=circle,draw=black,fill=red,text=black,inner sep = 0.1cm] (4tilde) at (1+\xspace,2) {$4$};
	\node[shape=rectangle,draw=black,fill=red,text=black,inner sep = 0.25cm] (b22tilde) at (2+\xspace,2) {};
	
	\node[shape=rectangle,draw=black,fill=blue,text=black,inner sep = 0.25cm] (b13tilde) at (-1+\xspace,1) {};
%	\node[shape=circle,draw=black,fill=blue,text=white,inner sep = 0.1cm] (5tilde) at (0+\xspace,1) {$5$};
	\node[shape=circle,draw=black,fill=blue,text=white,inner sep = 0.1cm] (3tilde) at (0.5+\xspace,1) {$3$};
	\node[shape=rectangle,draw=black,fill=red,text=black,inner sep = 0.25cm] (b23tilde) at (2+\xspace,1) {};

	\path (b11tilde) edge  (1tilde);
%	\path (1tilde) edge  (2tilde);
	\path (1tilde) edge  (b21tilde);
	\path (b12tilde) edge  (2tilde);
%	\path (3tilde) edge  (4tilde);
	\path (2tilde) edge  (b22tilde);
	\path (b13tilde) edge  (3tilde);
%	\path (5tilde) edge  (6tilde);
	\path (3tilde) edge  (b23tilde);
	
	%%------- Bottom Right

	\node at (0.5+\xspace,3.75-\yspace) {State $\tilde{s}^\prime$};

	\node[shape=rectangle,draw=black,fill=blue,text=black,inner sep = 0.25cm] (b11tildeprime) at (-1+\xspace,3-\yspace) {};
	\node[shape=circle,draw=black,fill=blue,text=white,inner sep = 0.1cm] (1tildeprime) at (0.5+\xspace,3-\yspace) {$1$};
%	\node[shape=circle,draw=black,fill=blue,text=white,inner sep = 0.1cm] (2tildeprime) at (1+\xspace,3-\yspace) {$2$};
	\node[shape=rectangle,draw=black,fill=red,text=black,inner sep = 0.25cm] (b21tildeprime) at (2+\xspace,3-\yspace) {};
	
	\node[shape=rectangle,draw=black,fill=blue,text=black,inner sep = 0.25cm] (b12tildeprime) at (-1+\xspace,2-\yspace) {};
	\node[shape=circle,draw=black,fill=red,text=black,inner sep = 0.1cm] (2tildeprime) at (0.5+\xspace,2-\yspace) {$2$};
%	\node[shape=circle,draw=black,fill=red,text=black,inner sep = 0.1cm] (4tildeprime) at (1+\xspace,2-\yspace) {$4$};
	\node[shape=rectangle,draw=black,fill=red,text=black,inner sep = 0.25cm] (b22tildeprime) at (2+\xspace,2-\yspace) {};
	
	\node[shape=rectangle,draw=black,fill=blue,text=black,inner sep = 0.25cm] (b13tildeprime) at (-1+\xspace,1-\yspace) {};
	\node[shape=circle,draw=black,fill=blue,text=white,inner sep = 0.1cm] (3tildeprime) at (0.5+\xspace,1-\yspace) {$3$};
%	\node[shape=circle,draw=black,fill=blue,text=white,inner sep = 0.1cm] (6tildeprime) at (1+\xspace,1-\yspace) {$6$};
	\node[shape=rectangle,draw=black,fill=red,text=black,inner sep = 0.25cm] (b23tildeprime) at (2+\xspace,1-\yspace) {};

	\path (b11tildeprime) edge  (1tildeprime);
%	\path (1tildeprime) edge  (2tildeprime);
	\path (1tildeprime) edge  (b21tildeprime);
	\path (b12tildeprime) edge  (2tildeprime);
%	\path (3tildeprime) edge  (4tildeprime);
	\path (2tildeprime) edge  (b22tildeprime);
	\path (b13tildeprime) edge  (3tildeprime);
%	\path (5tildeprime) edge  (6tildeprime);
	\path (3tildeprime) edge  (b23tildeprime);

	\path[->,>=stealth',shorten >=0pt,thick] (2.8,2) edge (\xspace-1.8,2);
	
	\path[->,>=stealth',shorten >=0pt,thick] (2.8,2-\yspace) edge (\xspace-1.8,2-\yspace);
	
%	\node at (-4,2.25) {\color{blue} Player 1};
%	\node at (5,2.25) {\color{red} Player 2};
	
%	\node (b1)at (-1.4,2.25) {\color{blue} $\mathcal{B}^1$};
%	\node (b2) at (2.4,2.25) {\color{red} $\mathcal{B}^2$};
	
%	\path[->,>=stealth',shorten >=1pt,thick,blue] (b1) edge  (b11);
%	\path[->,>=stealth',shorten >=1pt,thick,blue] (b1) edge  (b12);
%	
%	\path[->,>=stealth',shorten >=1pt,thick,red] (b2) edge  (b21);
%	\path[->,>=stealth',shorten >=1pt,thick,red] (b2) edge  (b22);

%	\node at (-2.5,3) {\color{blue} Commander 1};
%	\node at (3.5,3) {\color{red} Commander 1};
%	\node at (-2.5,1.5) {\color{blue} Commander 2};
%	\node at (3.5,1.5) {\color{red} Commander 2};

    \end{tikzpicture}
\caption{Counterexample to the isotonicity of the optimal value function without Assumption \ref{ass:rfc}.}
\label{fig:counter_example}
    \end{figure}
    
We also consider four states $s^1 = (1,1,1)$, $\tilde{s}^\prime = (1,2,1)$, $\tilde{s} = (2,2,1)$, $s^2=(2,2,2)$, for which we assume that all base attack success probabilities of uncontrolled objectives are 0:
\begin{align*}
\alpha^2_{1,s^1} = \alpha^2_{2,s^1} = \alpha^2_{3,s^1} = \alpha^2_{1,\tilde{s}^\prime} = \alpha^1_{2,\tilde{s}^\prime} = \alpha^2_{3,\tilde{s}^\prime} = \alpha^1_{1,\tilde{s}} =  \alpha^1_{2,\tilde{s}} =  \alpha^2_{3,\tilde{s}} =  \alpha^1_{1,s^2} = \alpha^1_{2,s^2} = \alpha^1_{3,s^2} = 0.
\end{align*}

As a result, $s^1$, $\tilde{s}^\prime$, $\tilde{s} $, $s^2$ are absorbing states that satisfy
\begin{align*}
V^*(s^1) = 0, \quad V^*(\tilde{s}^\prime) = \frac{\ell_2}{1-\gamma}, \quad V^*(\tilde{s}) = \frac{\ell_1+\ell_2}{1-\gamma}, \quad V^*(s^2) = \frac{\ell_1 + \ell_2 +\ell_3}{1-\gamma}.
\end{align*}

We note that Assumption \ref{ass:atk} holds, but Assumption \ref{ass:rfc} is violated since $\alpha^1_{1,s^\prime}+\alpha^2_{1,s} > 1$.

Let $a^{2^\dag} = (\textit{atk},\textit{atk},\textit{none}) \in \A^2_s$. 
We then obtain the following lower bound:
\begin{align*}
V^*(s) \geq \min_{a^1 \in \A^1_s} R(V^*,s,a^1,a^{2^\dag}) = L(s) + \gamma V^*(\tilde{s})  = \ell_3 + \gamma \frac{\ell_1 + \ell_2}{1-\gamma}.
\end{align*}

Similarly, let $a^{1^\dag} = (\textit{atk},\textit{none},\textit{atk}) \in \A^1_{s^\prime}$. 
We then obtain the following upper bound:
\begin{align*}
V^*(s^\prime) \leq \max_{a^2 \in \A^2_{s^\prime}} R(V^*,s^\prime,a^{1^\dag},a^{2}) = L(s^\prime) + \gamma V^*(\tilde{s}^\prime)  = \ell_1 + \ell_3 + \gamma \frac{\ell_2}{1-\gamma}.
\end{align*}

Since $\ell_2 = \ell_3$ and $\ell_1 + \ell_2 > \ell_3$, we then obtain the non-isotonicity of the optimal value function  when $\gamma > 0.5$:
\begin{align*}
V^*(s) \geq \frac{(1-\gamma)\cdot\ell_3 + \gamma \cdot(\ell_1 + \ell_2)}{1-\gamma}  > \frac{(1-\gamma)\cdot(\ell_1 + \ell_3) + \gamma \cdot\ell_2}{1-\gamma}  \geq V^*(s^\prime).
\end{align*}

\hfill$\triangle$
\end{counterexample}

We next investigate the states that are achievable and the actions that are selected with positive probability in equilibrium of the stochastic game $\Gamma$. To this end, we  propose the following axis classification at a given state and define \emph{battle fronts} for each player. Consider an axis $x = (o_1,\dots,o_{n}) \in \X$ of size $n$ and a state $s \in \S$.

\begin{henumerate}
\item If for every $k \in \{1,\dots,n\}$, $s_{o_k} = 1$ (i.e., every objective on that axis is controlled by Player $1$), then we say that $x$ is of type $\type{x}{s} = \textit{c1}$, and the battle front occurs at the objective $\front{i}{x}{s} = \{o_n\}$ for each player $i$.

\item If for every $k \in \{1,\dots,n\}$, $s_{o_k} = 2$ (i.e., every objective on that axis is controlled by Player $2$), then we say that $x$ is of type $\type{x}{s} = \textit{c2}$, and the battle front occurs at the objective $\front{i}{x}{s} = \{o_1\}$ for each player $i$. 

\item If there exists $k \in \{1,\dots,n-1\}$ such that $s_{o_1} = \cdots = s_{o_k} = 1$ and $s_{o_{k+1}} = \cdots = s_{o_n} =2$, then we say that $x$ is of type $\type{x}{s} = \textit{pf}$, and a pure front exists at the objectives $\front{i}{x}{s} = \{o_k,o_{k+1}\}$ for each player $i$.

\item If there exists $k \in \{1,\dots,n-1\}$ such that $s_{o_1} = \cdots = s_{o_{k-1}} = 1$, $s_{o_{k}}=2$, $s_{o_{k+1}}=1$,  and $s_{o_{k+2}} = \cdots = s_{o_n} =2$, then we say that $x$ is of type $\type{x}{s} = \textit{sf}$, and a split front exists, i.e., a front exists for Player 1 at objective $\front{1}{x}{s} = \{o_{k}\}$ and another front exists for Player 2 at objective $\front{2}{x}{s} = \{o_{k+1}\}$.

\end{henumerate}

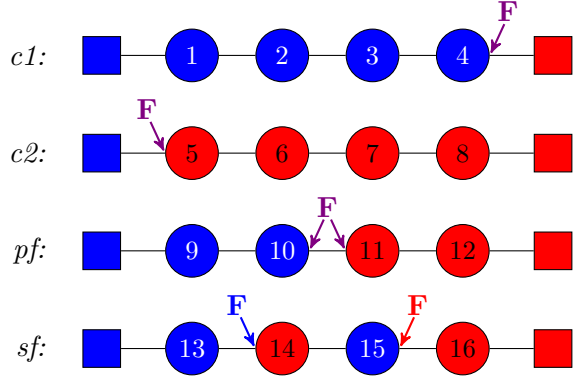
\begin{figure}[htbp]
\centering
        \small
        \begin{tikzpicture}[x = 1.2cm,y=1.3cm]
%        \node[text=black](d1) at (-0.7,2.3) {$0.6$};

	\node[shape=rectangle,draw=black,fill=blue,text=black,inner sep = 0.25cm] (b11) at (-1,3) {};
	\node[shape=circle,draw=black,fill=blue,text=white,inner sep = 0.1cm, minimum size=0.7cm] (1) at (0,3) {$1$};
	\node[shape=circle,draw=black,fill=blue,text=white,inner sep = 0.1cm, minimum size=0.7cm] (2) at (1,3) {$2$};
	\node[shape=circle,draw=black,fill=blue,text=white,inner sep = 0.1cm, minimum size=0.7cm] (3) at (2,3) {$3$};
	\node[shape=circle,draw=black,fill=blue,text=white,inner sep = 0.1cm, minimum size=0.7cm] (4) at (3,3) {$4$};
	\node[shape=rectangle,draw=black,fill=red,text=black,inner sep = 0.25cm] (b21) at (4,3) {};
	
	\node[shape=rectangle,draw=black,fill=blue,text=black,inner sep = 0.25cm] (b12) at (-1,2) {};
	\node[shape=circle,draw=black,fill=red,text=black,inner sep = 0.1cm, minimum size=0.7cm] (5) at (0,2) {$5$};
	\node[shape=circle,draw=black,fill=red,text=black,inner sep = 0.1cm, minimum size=0.7cm] (6) at (1,2) {$6$};
	\node[shape=circle,draw=black,fill=red,text=black,inner sep = 0.1cm, minimum size=0.7cm] (7) at (2,2) {$7$};
	\node[shape=circle,draw=black,fill=red,text=black,inner sep = 0.1cm, minimum size=0.7cm] (8) at (3,2) {$8$};
	\node[shape=rectangle,draw=black,fill=red,text=black,inner sep = 0.25cm] (b22) at (4,2) {};
	
	\node[shape=rectangle,draw=black,fill=blue,text=black,inner sep = 0.25cm] (b13) at (-1,1) {};
	\node[shape=circle,draw=black,fill=blue,text=white,inner sep = 0.1cm, minimum size=0.7cm] (9) at (0,1) {$9$};
	\node[shape=circle,draw=black,fill=blue,text=white,inner sep = 0.1cm, minimum size=0.7cm] (10) at (1,1) {$10$};
	\node[shape=circle,draw=black,fill=red,text=black,inner sep = 0.1cm, minimum size=0.7cm] (11) at (2,1) {$11$};
	\node[shape=circle,draw=black,fill=red,text=black,inner sep = 0.1cm, minimum size=0.7cm] (12) at (3,1) {$12$};
	\node[shape=rectangle,draw=black,fill=red,text=black,inner sep = 0.25cm] (b23) at (4,1) {};
	
	\node[shape=rectangle,draw=black,fill=blue,text=black,inner sep = 0.25cm] (b14) at (-1,0) {};
	\node[shape=circle,draw=black,fill=blue,text=white,inner sep = 0.1cm, minimum size=0.7cm] (13) at (0,0) {$13$};
	\node[shape=circle,draw=black,fill=red,text=black,inner sep = 0.1cm, minimum size=0.7cm] (14) at (1,0) {$14$};
	\node[shape=circle,draw=black,fill=blue,text=white,inner sep = 0.1cm, minimum size=0.7cm] (15) at (2,0) {$15$};
	\node[shape=circle,draw=black,fill=red,text=black,inner sep = 0.1cm, minimum size=0.7cm] (16) at (3,0) {$16$};
	\node[shape=rectangle,draw=black,fill=red,text=black,inner sep = 0.25cm] (b24) at (4,0) {};

	\path (b11) edge (1);
	\path (1) edge  (2);
	\path (2) edge  (3);
	\path (3) edge  (4);
	\path (4) edge  (b21);
	\path (b12) edge (5);
	\path (5) edge  (6);
	\path (6) edge  (7);
	\path (7) edge  (8);
	\path (8) edge  (b22);
	\path (b13) edge (9);
	\path (9) edge  (10);
	\path (10) edge  (11);
	\path (11) edge  (12);
	\path (12) edge  (b23);
	\path (b14) edge (13);
	\path (13) edge  (14);
	\path (14) edge  (15);
	\path (15) edge  (16);
	\path (16) edge  (b24);
	
	\node[anchor=east] at (-1.5,3) {\normalsize\textit{c1:}};
	\node[anchor=east] at (-1.5,2) {\normalsize\textit{c2:}};
	\node[anchor=east] at (-1.5,1) {\normalsize\textit{pf:}};
	\node[anchor=east] at (-1.5,0) {\normalsize\textit{sf:}};
	
	\node at (3.5,3.45) {\normalsize\color{violet}\textbf{F}};
	\draw[->,>=stealth',shorten >=1pt,thick,violet] (3.46,3.32) -- (3.3,3);
	
	\node at (0.5-1,2.45) {\normalsize\color{violet}\textbf{F}};
	\draw[->,>=stealth',shorten >=1pt,thick,violet] (0.54-1,2.32) -- (0.7-1,2);
	
	\node at (1.5,1.45) {\normalsize\color{violet}\textbf{F}};
	\draw[->,>=stealth',shorten >=1pt,thick,violet] (1.54,1.32) -- (1.7,1);
	\draw[->,>=stealth',shorten >=1pt,thick,violet] (1.46,1.32) -- (1.3,1);
	
	\node at (0.5,0.45) {\normalsize\color{blue}\textbf{F}};
	\node at (2.5,0.45) {\normalsize\color{red}\textbf{F}};
	
	\draw[->,>=stealth',shorten >=1pt,thick,blue] (0.54,0.32) -- (0.7,0);
	\draw[->,>=stealth',shorten >=1pt,thick,red] (2.46,0.32) -- (2.3,0);

%	\node at (-4,2.25) {\color{blue} Player 1};
%	\node at (5,2.25) {\color{red} Player 2};
%	
%	\node at (-2.5,3) {\color{blue} Commander 1};
%	\node at (3.5,3) {\color{red} Commander 1};
%	\node at (-2.5,1.5) {\color{blue} Commander 2};
%	\node at (3.5,1.5) {\color{red} Commander 2};
%	\draw (0,3) node[cross=7.5pt, blue, line width = 0.5mm,scale=0.15] {};    
		
    \end{tikzpicture}
     \caption{Axis classification: In \textit{c1}, the front (denoted F) for both players is at objective 4. In \textit{c2} the front for both players is at objective 5. In \textit{pf} the front for both players is at objectives 10 and 11. Finally, for \textit{sf} the front is at objective 14 for Player 1, and at objective 15 for Player 2.}
    \label{fig:axis_states}
\end{figure}

\noindent We illustrate these axis types and the battle fronts in Figure \ref{fig:axis_states}. 

We make the following assumption regarding the initial state of the campaign:

\begin{assumption}\label{ass:init}
    At the initial state $s^{(0)} \in \S$, every axis $x \in \X$ is of type $\type{x}{s^{(0)}} \in \{\textit{c1},\textit{c2},\textit{pf},\textit{sf}\}$.
\end{assumption}

A military campaign postulates opposing forces that are initially separated by geographic borders or necessary campaign milestones (e.g., a force cannot gain air parity and then air superiority without suppressing or neutralizing opponent air defenses first). Assumption \ref{ass:init} and the initial distribution of the campaign extend from the players initial disposition to how an opponent succeeds at the immediate outset of an attack. Thus, we assume that a force cannot seize objectives past their supply chains or skip necessary campaign milestones before achieving subsequent objectives. 

In the next proposition, we leverage the game's structure and Theorem \ref{thm:isotone} to derive properties of the states and selected actions in equilibrium.

\begin{proposition}\label{prop:reduce}
The set of achievable states for any policy profile is
\begin{align}
\{s \in \{1,2\}^{\O} \ | \ \type{x}{s} \in \{\textit{c1},\textit{c2},\textit{pf},\textit{sf}\}, \ \forall x \in \X\}.\label{new:state}
\end{align}

There exists an MPE such that for every achievable state $s \in \S$, each player $i\in\{1,2\}$ randomizes over actions $a^i \in \A^i_s$ satisfying
\begin{align}
&\forall x \in \X, \ \forall o \in x, \ a_{o}^i \in \begin{cases} \{\textit{none},\textit{atk}\} & \text{if } o \in \front{i}{x}{s} \text{ and } s_o=-i \\\{\textit{none},\textit{rfc}\} & \text{if } o \in \front{i}{x}{s} \text{ and } s_o=i  \\ \{\textit{none}\} & \text{otherwise}\end{cases}\label{new_action_1}\\
&\forall c \in \C^i, \ |\{a^i_o \in \{\textit{atk},\textit{rfc}\}, \ \forall o \in \O_c\}| = 1.\label{new_action_2}
\end{align}
Furthermore, if the game parameters satisfy
\begin{align}
\ell_o > 0, \quad &\forall o \in \O\label{cond1}\\
\alpha^i_{o,s} > 0, \quad & \forall i \in \{1,2\}, \ \forall s \in \S, \ \forall o \in \overline{\O}^i_s \ | \  s_o = -i\\
0<\rho^i_{o,s} < 1, \quad & \forall i \in \{1,2\}, \ \forall s \in \S, \ \forall o \in \overline{\O}^i_s \ | \  s_o = i,\label{cond3}
\end{align}
then the actions selected with positive probability in \emph{every} MPE satisfy \eqref{new_action_1}-\eqref{new_action_2}.

\end{proposition}

From this proposition, we obtain that under Assumption \ref{ass:init}, each axis at each stage of the campaign will belong to one of the four defined types. As a result, we can reduce the state space $\S$ to the set of achievable states \eqref{new:state}, which is of size $\prod_{x \in \X}2|x|$. This provides a substantial memory and runtime improvement: For a campaign instance with 5 axes containing 5 objectives each, the state space will be reduced by 99.7\% (from 33.55 million to 100 thousand states).

Using game-theoretic arguments and the isotonocity of the optimal value function (Theorem \ref{thm:isotone}), we also obtain properties of the players' mixed strategies implemented at each achievable state in equilibrium. At any state $s \in \S$, each player has an incentive to order each commander to reinforce or attack one objective. We also naturally observe that if a commander attacks or reinforces an objective in an axis $x \in \X$, their incentive is to do so at the battle front. If $\type{x}{s} = \textit{c1}$ (resp. \textit{c2}) then Player 1's (resp. Player 2's) incentive is to reinforce the last (resp. first) objective of the axis, while Player 2's (resp. Player 1's) incentive is to attack it. If $\type{x}{s} =  \textit{pf}$, then each player may either reinforce the objective at the front they control, or attack the objective controlled by their opponent. If each player decides to attack the objective they do not control, and each resulting battle succeeds, then the axis's new type becomes $\textit{sf}$, as a split front arises. In any \textit{sf} situation, players do not have an open LoC to their farthest controlled objective and are unable to reinforce that objective or attack farther objectives. Accordingly, each player $i$ only has an incentive to attack their closest uncontrolled objective $\front{i}{x}{s}$.

Thus, we can reduce the players' action spaces by removing the actions that do not satisfy \eqref{new_action_1}-\eqref{new_action_2}. Proposition \ref{prop:reduce} shows that in almost all cases, this operation does not remove any MPE of $\Gamma$, while in some edge cases (i.e., when the game parameters do not satisfy \eqref{cond1}-\eqref{cond3}), at least one MPE will be retained. This also provides a computational gain, as the action space sizes are reduced to $\prod_{c \in \C^i}\sum_{x \in \X_c}|\front{i}{x}{s}| $ for every player $i \in \{1,2\}$ and every achievable state $s \in \S$.  
Henceforth, we assume that the state space and the players' action spaces are respectively reduced to \eqref{new:state} and \eqref{new_action_1}-\eqref{new_action_2}. 

\subsection{Accelerated Shapley's Value Iteration}

We propose to accelerate the VI algorithm described in Section \ref{sec:VI} by reducing the number or size of linear programs to solve. Indeed, the campaign's structure leads to equilibrium properties of the zero-sum matrix game solved at each iteration of the algorithm. In the next proposition, we analyze the matrix game when each player has one commander.

\begin{proposition}\label{prop:reduction}
Consider a campaign with one commander for each player. Let $t \in \mathbb{Z}_{>0}$ be an iteration of the VI algorithm, and $s \in \S$ be a state.
\begin{itemize}
\item[--] If $|\X| = 1$, then the zero-sum matrix game $\Gamma(V^{(t-1)},s)$ admits a pure equilibrium.

\item[--] If at least one axis $x \in \X$ is of type $\type{x}{s} = \textit{pf}$, then at least one of the players does not reinforce any objective in axis $x$ in at least one equilibrium of $\Gamma(V^{(t-1)},s)$.

\end{itemize}

\end{proposition}

By leveraging the game's structure and the isotonicity result of the function $V^{(t-1)}$ (showed in the proof of Theorem \ref{thm:isotone}), we were able to analytically characterize pure equilibria of the matrix game solved at each iteration of the VI algorithm when there is a single axis. Furthermore, when there are multiple axes under the responsibility of a single commander, Proposition \ref{prop:reduction} shows that the action sets can be further reduced before solving the matrix game.

This proposition suggests that in general, it is computationally valuable to search for pure equilibria and eliminate actions before attempting to solve the game via linear programing. This leads us to implement the following algorithmic steps:
At every iteration $t \in \mathbb{Z}_{> 0}$ and every state $s \in \S$, we first determine whether $\Gamma(V^{(t-1)},s)$ admits an equilibrium in pure strategies, which occurs when
\begin{align*}
\min_{a^1 \in \A^1_s} \max_{a^2 \in \A^2_s} R(V^{(t-1)},s,a^1,a^2) = \max_{a^2 \in \A^2_s} \min_{a^1 \in \A^1_s}  R(V^{(t-1)},s,a^1,a^2).
\end{align*}
If the equality holds, then $(\pi^{1^*}(s),\pi^{2^*}(s)) \in \Delta(\A^1_s)\times\Delta(\A^2_s)$ defined by $\pi^{1^*}(s,a^{1^*}) = 1$, $\pi^{2^*}(s,a^{2^*}) = 1$ where
\begin{align*}
a^{1^*} \in \argmin_{a^1 \in \A^1_s} \max_{a^2 \in \A^2_s} R(V^{(t-1)},s,a^1,a^2), \quad \quad \quad a^{2^*} \in \argmax_{a^2 \in \A^2_s}\min_{a^1 \in \A^1_s} R(V^{(t-1)},s,a^1,a^2),
\end{align*}
form a pure equilibrium of $\Gamma(V^{(t-1)},s)$ and can be efficiently computed by searching over all values of the matrix $R(V^{(t-1)},s)$.

If a pure equilibrium does not exist, we next seek to simplify the game $\Gamma(V^{(t-1)},s)$: We implement an iterated elimination of weakly dominated strategies to remove actions while retaining at least one equilibrium of $\Gamma(V^{(t-1)},s)$. For Player 1, we say that action $a^1 \in \A^1_s$ \emph{dominates} action $a^{1^\prime} \in \A^1_s$ in the game $\Gamma(V^{(t-1)},s)$ if $R(V^{(t-1)},s,a^1,a^2) \leq R(V^{(t-1)},s,a^{1^\prime},a^2)$ for every $a^2 \in \A^2_s$. Similarly, we say that Player 2's action $a^2 \in \A^2_s$ dominates action $a^{2^\prime} \in \A^2_s$ in the game $\Gamma(V^{(t-1)},s)$ if $R(V^{(t-1)},s,a^1,a^2) \geq R(V^{(t-1)},s,a^1,a^{2^\prime})$ for every $a^1 \in \A^1_s$. 

This algorithm step iteratively removes weakly dominated actions, and stops when no action is removed for either player, producing smaller sets of actions $\bar{\A}^1_s \subseteq \A^1_s$ and $\bar{\A}^2_s \subseteq \A^2_s$. Once the iterated elimination terminates, the algorithm solves the smaller linear program $\operatorname{LP}(R(V^{(t-1)},s),\bar{\A}^1_s,\bar{\A}^2_s)$. These modifications to solve zero-sum matrix games are described in Algorithm \ref{alg:AZS} and are embedded into the VI algorithm, resulting in an \emph{accelerated value iteration} (AVI) algorithm, summarized in Algorithm \ref{alg:AVI}. 

\begin{algorithm}[htbp]
\caption{Accelerated Zero-Sum Matrix Game Solution (AZS($s,\A^{1}_s,\A^{2}_s,R$))} \label{alg:AZS}
\normalsize
\SetArgSty{textnormal}
\SetKwInOut{Parameters}{Parameters}
\SetKwInOut{Input}{Input}
\SetKwInOut{Output}{Output}
\SetKwIF{nlIf}{ElseIf}{Else}{if}{then}{else if}{else}{\nl}
\SetKwRepeat{Do}{do}{while}
%\Parameters{Integers $k_1$, $k_2$, and $b$, positive number $\varepsilon$}
\Input{State $s$, action spaces $\A^{1}_s$ and $\A^{2}_s$, and payoff matrix $R \in \mathbb{R}^{\A^{1}_s\times \A^{2}_s}$}

%Game $\Gamma = \langle\S,(\A^1,\A^2),P,L,\gamma\rangle$ and optimality gap $\epsilon \in \mathbb{R}_{>0}$
\Output{Optimal value $V^\prime(s)$ and equilibrium $(\pi^{1^\prime}(s),\pi^{2^\prime}(s)) \in \Delta(\A^{1^\prime}_s)\times \Delta(\A^{2^\prime}_s)$ of the matrix game}
\BlankLine
%\nl Initialize: Reduced Cost $ (\Bar{c})_{p \in \mathcal{P}} \gets \, -M , \quad $ Parameter $(\kappa)_{p \in \mathcal{P}} \gets 0$ \;
%\nl Path set as warm-start $\tilde{\Lambda} \gets \tilde{\Lambda} \setminus \{\lambda \in \tilde{\Lambda} |  i \in \lambda \cap \mathcal{Y}^{0} \}$  and define $\mathcal{L}_{path}(\mathcal{Y}^{0},\mathcal{Y}^{1},\tilde{\Lambda})$ \md{Do this when assigning to child node} \;

%$t \gets 0$\;
%
%$V^{(t)} \gets L$\;

%\Do{$\| V^{(t)} - V^{(t-1)}\|_{\infty} > \epsilon (1-\gamma)/(2\gamma)$}{
%$t \gets t+1$\;
%
%\For{every $s \in \S$}{

%$R(V^{(t-1)},s,a^1,a^2) \gets L(s) + \gamma  \cdot\sum_{s^\prime \in \S} P(s^\prime \, | \, s, a^{1},a^{2}) \cdot V^{(t-1)}(s^\prime), \ \forall (a^1,a^2) \in \A^1_s \times \A^2_s$\;

Compute $a^{1^\prime}\in \arg \min\limits_{a^1 \in \A^{1}_s} \max\limits_{a^2 \in \A^{2}_s} R(a^1,a^2)$ and $a^{2^\prime} \in \arg \max\limits_{a^2 \in \A^{2}_s}  \min\limits_{a^1 \in \A^{1}_s} R(a^1,a^2)$\;

 \If{$\max\limits_{a^2 \in \A^{2}_s} R(a^{1^\prime},a^2) =  \min\limits_{a^1 \in \A^{1}_s}R(a^1,a^{2^\prime})$}{
 
 $\pi^{i^\prime}(s,a^{i^\prime}) \gets 1$ \ and \ $\pi^{i^\prime}(s,a^i) \gets 0$,  $\forall i \in \{1,2\}$, $\forall a^i \in \A^{i}_s\setminus\{a^{i^\prime}\}$\;
 
 $V^{\prime}(s) \gets R(a^{1^\prime},a^{2^\prime})$\;
 
 }
 \Else{
 
 $\bar{\A}^{i}_s \gets \A^{i}_s$, $\forall i \in \{1,2\}$\;
 
 \Do{$Reduced = False$}
 {
$Reduced \gets True$\;
 
 \For{every $i \in \{1,2\}$ and every $a^i \neq a^{i^\dag} \in \bar{\A}^i_s$}{
% 
% $(a^i,a^{i^\prime}) \in (\bar{\A}^_s)^2 \times \A^{i^\prime}_s$ such that $a^i \neq a^{i^\prime}$
% }{
 \If{$a^i$ dominates $a^{i^\dag}$}
 {
 $\bar{\A}^i_s \gets \bar{\A}^i_s\setminus \{a^{i^\dag}\}$\;
% Remove the row/column corresponding to $a^{i^\prime}$ from $R(V^{(t-1)},s)$\;
 
% Remove $a^{i^\prime}$ from $A^{i^\prime}_s$\;
 
 $Reduced \gets False$\;
 
 }
 }
 }
%Iterated Elimination of Dominated Strategies\;

Solve $\operatorname{LP}(R,\bar{\A}^1_s,\bar{\A}^2_s))$\; 

\hspace{0.5cm} $V^{\prime}(s) \gets$ optimal value, \ $(\pi^{1^\prime}(s),\pi^{2^\prime}(s)) \gets $ optimal primal and dual solutions\; 
 }

\nl \KwRet{$V^{\prime}(s)$, $(\pi^{1^\prime}(s),\pi^{2^\prime}(s))$ }
\end{algorithm}

\begin{algorithm}[htbp]
\caption{Accelerated Value Iteration (AVI)} \label{alg:AVI}
\normalsize
\SetArgSty{textnormal}
\SetKwInOut{Parameters}{Parameters}
\SetKwInOut{Input}{Input}
\SetKwInOut{Output}{Output}
\SetKwIF{nlIf}{ElseIf}{Else}{if}{then}{else if}{else}{\nl}
\SetKwRepeat{Do}{do}{while}
%\Parameters{Integers $k_1$, $k_2$, and $b$, positive number $\varepsilon$}
\Input{Game $\Gamma = \langle\S,(\A^1,\A^2),P,L,\gamma\rangle$ and optimality gap $\epsilon \in \mathbb{R}_{>0}$}
\Output{$\epsilon/2-$approximate game value $V^{(t)} \in \mathbb{R}^\S$ and $\epsilon$-MPE $(\pi^{1^\prime},\pi^{2^\prime}) \in \Delta^1\times \Delta^2$}
\BlankLine
%\nl Initialize: Reduced Cost $ (\Bar{c})_{p \in \mathcal{P}} \gets \, -M , \quad $ Parameter $(\kappa)_{p \in \mathcal{P}} \gets 0$ \;
%\nl Path set as warm-start $\tilde{\Lambda} \gets \tilde{\Lambda} \setminus \{\lambda \in \tilde{\Lambda} |  i \in \lambda \cap \mathcal{Y}^{0} \}$  and define $\mathcal{L}_{path}(\mathcal{Y}^{0},\mathcal{Y}^{1},\tilde{\Lambda})$ \md{Do this when assigning to child node} \;

$t \gets 0$\;

$V^{(t)} \gets L$\;

\Do{$\| V^{(t)} - V^{(t-1)}\|_{\infty} > \epsilon (1-\gamma)/(2\gamma)$}{
$t \gets t+1$\;

\For{every $s \in \S$}{

$R(V^{(t-1)},s,a^1,a^2) \gets L(s) + \gamma  \cdot\sum_{s^\prime \in \S} P(s^\prime \, | \, s, a^{1},a^{2}) \cdot V^{(t-1)}(s^\prime), \ \forall (a^1,a^2) \in \A^1_s \times \A^2_s$\;

 $(V^{(t)}(s),(\pi^{1^\prime}(s),\pi^{2^\prime}(s))) \gets $ AZS($s,\A^1_s,\A^2_s,R(V^{(t-1)},s)$)\; 
 
% 
%Solve $a^{1^*} \hspace{-0.1cm}\in \arg \min\limits_{a^1 \in \A^1_s} \max\limits_{a^2 \in \A^2_s} \hspace{-0.05cm}R(V^{(t-1)},s,a^1,a^2)$ and $a^{2^*} \hspace{-0.1cm}\in \arg \max\limits_{a^2 \in \A^2_s}  \min\limits_{a^1 \in \A^1_s} \hspace{-0.05cm}R(V^{(t-1)},s,a^1,a^2)$\;
%
% \If{$\max\limits_{a^2 \in \A^2_s} R(V^{(t-1)},s,a^{1^*},a^2) =  \min\limits_{a^1 \in \A^1_s}R(V^{(t-1)},s,a^1,a^{2^*})$}{
% 
% $\pi^{i^\prime}(s,a^{i^*}) \gets 1$ \ and \ $\pi^{i^\prime}(s,a^i) \gets 0$,  $\forall i \in \{1,2\}$, $\forall a^i \in \A^i_s\setminus\{a^{i^*}_s\}$\;
% 
% $V^{(t)}(s) \gets R(V^{(t-1)},s,a^{1^*},a^{2^*})$\;
% 
% }
% \Else{
% 
% $\bar{\A}^{i}_s \gets \A^i_s$, $\forall i \in \{1,2\}$\;
% 
% \Do{$Reduced = False$}
% {
%$Reduced \gets True$\;
% 
% \For{every $i \in \{1,2\}$ and every $a^i \neq a^{i^\prime} \in \bar{\A}^i_s$}{
%% 
%% $(a^i,a^{i^\prime}) \in (\bar{\A}^_s)^2 \times \A^{i^\prime}_s$ such that $a^i \neq a^{i^\prime}$
%% }{
% \If{$a^i$ dominates $a^{i^\prime}$}
% {
% $\bar{\A}^i_s \gets \bar{\A}^i_s\setminus \{a^{i^\prime}\}$\;
%% Remove the row/column corresponding to $a^{i^\prime}$ from $R(V^{(t-1)},s)$\;
% 
%% Remove $a^{i^\prime}$ from $A^{i^\prime}_s$\;
% 
% $Reduced \gets False$\;
% 
% }
% }
% }
%%Iterated Elimination of Dominated Strategies\;
%
%
%
%Solve $\operatorname{LP}(V^{(t-1)},s,\bar{\A}^1_s,\bar{\A}^2_s))$\; 
%
%\hspace{0.5cm} $V^{(t)}(s) \gets$ optimal value, \ $(\pi^{1^\prime}(s),\pi^{2^\prime}(s)) \gets $ optimal primal and dual solutions\; 
% }

}}
\nl \KwRet{$V^{(t)}$, $(\pi^{1^\prime},\pi^{2^\prime})$ }
\end{algorithm}

In the next section, we test the VI and AVI algorithms on representative campaign instances and show the computational gain provided by the equilibrium properties derived in this section.

%\subsection{Heuristic}

%As a consequence, the algorithm first arbitrarily selects a vector $V^{(0)} = (V^{(0)}(s))_{s \in \S}$ and iteratively applies the mapping $T$ by solving a zero-sum matrix game at every stage $s \in \S$, where the matrix payoff $R^(s,V) = (R_{a^1,a^2}(s,V))_{(a^1,a^2) \in \A^1_s \times \A^2_s}$ is given by:
%\begin{align*}
%\forall (a^1,a^2) \in \A^1_s \times \A^2_s, \ 
%\end{align*}

%
%First recall SVI (maybe in the text)
%
%\begin{align*}
%\forall s \in \S, \ V^*(s) = \min_{\pi^1(s) \in \Delta(\A^1_s)} \max_{a^2 \in \A^2_s} \left(L(s) + \gamma \sum_{a^1\sim\pi^1(s)} \pi^1(s,a^1) \sum_{s^\prime \in \S} P(s^\prime \ | \ s, a^{1},a^{2}) \cdot V^*(s^\prime) \right).
%\end{align*}

%axis types will belong in  will belong 

%We recall that the 
%$|\S| = 2^{|\O|}$ and $|\A^i_s| = \prod_{c \in \C} (1 + |\overline{\O}^i_s \cap \X_c|)$
%To mitigate the game's curse of dimensionality, 

%!TEX root = NRL Manuscript.tex

\section{Case Study} \label{sec:case_study}

This section illustrates our stochastic game application to military campaigns. We build our case study upon a fictional geopolitical scenario and derive insights from the MPE policy and value. We then compare the computational performance of the VI and AVI algorithms for military campaigns of various sizes.
All algorithms are implemented in Python v.3.8.5 and all optimization problems are solved using Gurobi v.10.0.1 on an Intel i7-9750H processor with a 2.60 GHz (1 core) CPU on a single thread with 16 GB of RAM.

\subsection{Campaign Geography}

We consider a fictitious country, Isthia, which is an isthmus under attack from an aggressor country, Adversary, to their East. Isthia controls a key canal that bisects the country. Recent geopolitical affairs have inhibited Adversary's use of this canal, so they seek to gain control of the canal and seize a portion of Isthia. Furthermore, Adversary seeks to threaten Isthia's sovereignty by seizing their seat of government and controlling their largest city. Controlling a large portion of Isthia will force favorable negotiations for Adversary. 

Isthia has a large mountain range through the East-West spine of the country that largely inhibits North-South travel for large military forces. There is a key corridor between and a road system adjacent to the canal, but other mountain roads cannot support large military forces. Isthia is surrounded on the North by the North Ocean and on the South by the South Ocean. The countries to the West of Isthia are allies and guarantee free maneuver and travel to coalition forces allied with Isthia. We refer to the coalition (Isthia and allies) as Player 1 and to Adversary as Player 2. 

\begin{figure}[htbp]
\centering
\includegraphics[scale=.65]{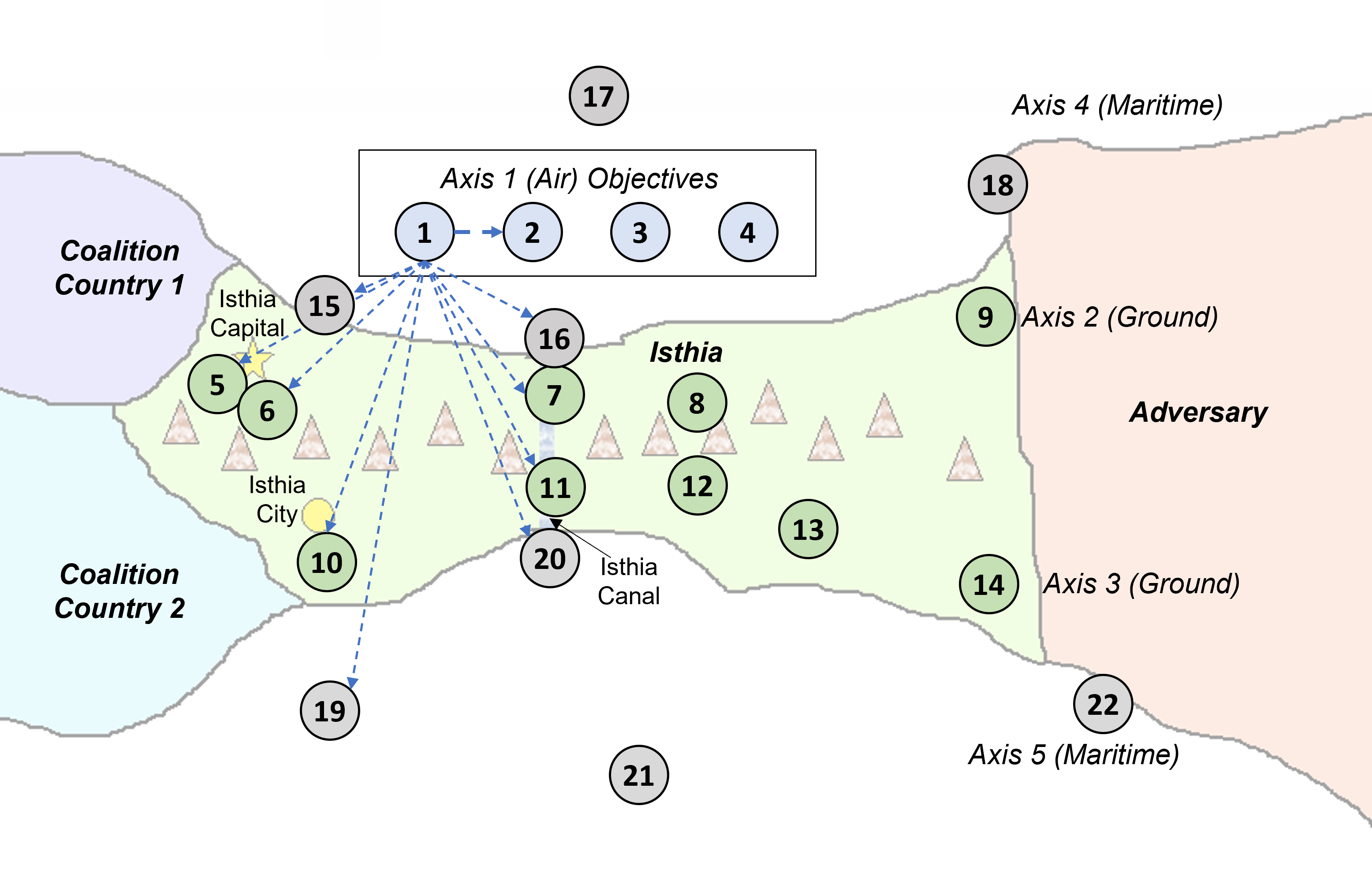}
\caption{Campaign geography and objectives. We illustrate the effect of Player 1 controlling \textit{Integrated Air Defense System} (obj. 1): Controlling obj. 1 is needed to successfully achieve \textit{Airspace 1} (obj. 2) but also increases the likelihood of successfully seizing objectives 5,6,7,10,11,15,16,19, and 20.}
\label{fig:fiction}
\end{figure}

There are three distinct commanders for each player: The air commander who is responsible for all airspace in the campaign (air axis), the ground commander who is responsible for the Northern axis and Southern axis (ground axes), and the maritime commander who is responsible for maritime operations in the North and South Oceans (maritime axes). We depict the case study geography and objectives in Figure \ref{fig:fiction}. The air objectives are either spread out geospatially---\textit{Integrated Air Defense System 1} and \textit{2} (objs. 1 and 4)---or they are not geographic---\textit{Airspace 1} and \textit{2} (objs. 2 and 3).

\begin{figure}[htbp]
\centering
\includegraphics[scale=.5]{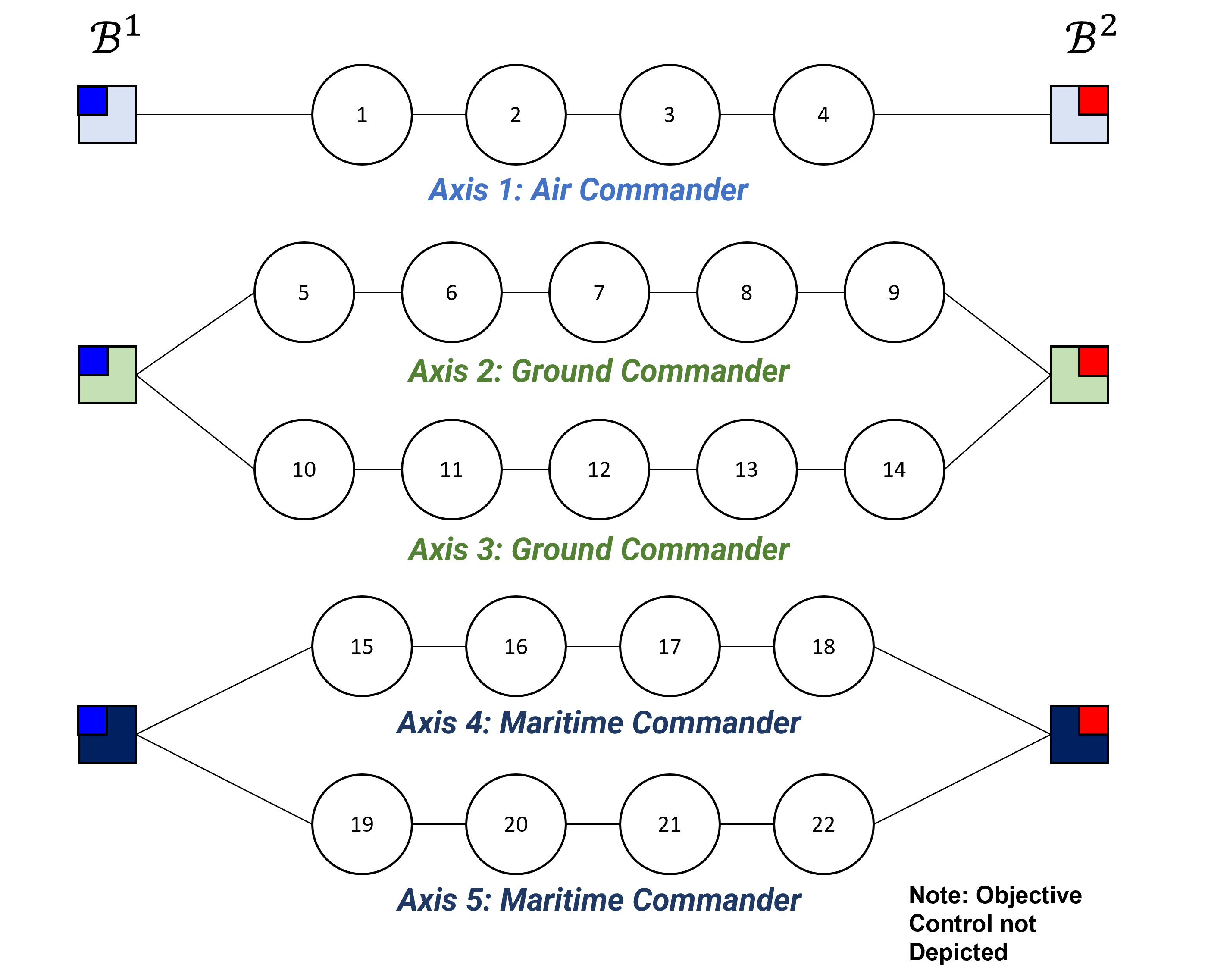}
\caption{Campaign axes and commanders}
\label{fig:22}
\end{figure}

\begin{figure}[htbp]
\centering
\includegraphics[scale=.6]{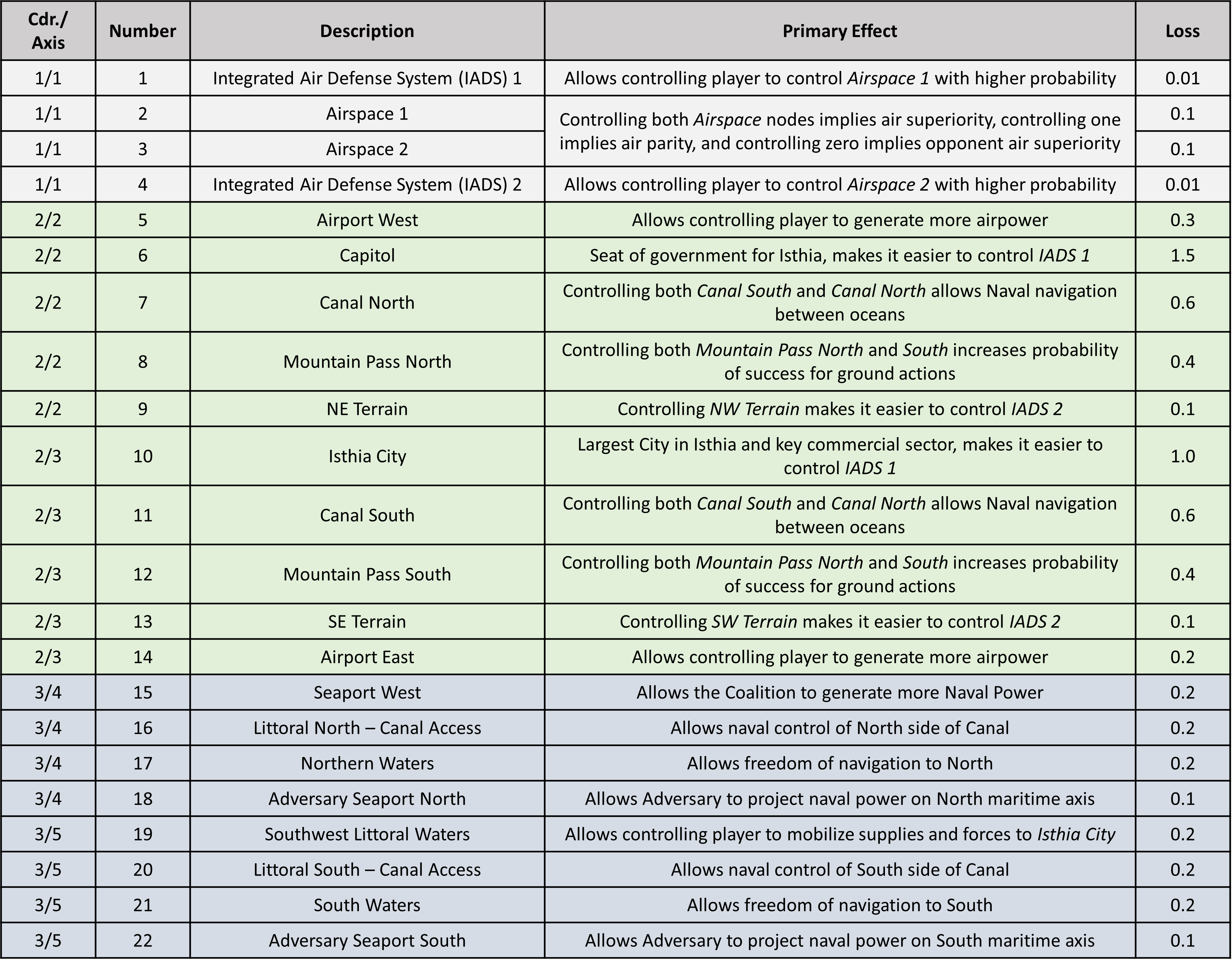}
\caption{Campaign objectives}
\label{fig:obj22}
\end{figure}

Figure \ref{fig:22} illustrates the axis structure for the 22-objective ($|\O|=22$), five-axis ($|\X|=5$) campaign. There are three commanders for each player: $|\C^1| = |\C^2| =3$, where $|\X_1|=1$, $|\X_2|=2$, and $|\X_3|=2$. To represent the interconnected nature of warfare, axes are connected probabilistically. For instance, controlling an airspace objective will make it more likely to succeed during an attack on a subset of ground axis objectives. Controlling both air objectives indicates \textit{air superiority} and will allow close air support (CAS), enabling higher likelihood of ground attack success. Similarly, controlling certain naval objectives will enable ground operations to succeed with higher probability. Controlling seaports and airports allows a player to generate more naval power or air power, which is reflected by higher probability of succeeding for actions in the naval and air axis respectively. Controlling the canal on ground allows naval forces to move between oceans, as reflected by increased likelihood of naval operational success. Similarly, controlling the mountain pass increases the likelihood of ground operational success on both axes. In Figure \ref{fig:obj22} we provide the objective losses and highlight one campaign effect for each objective.

\subsection{Base Attack and Reinforce Success Probabilities} \label{sec:soliciting_T}

In Section \ref{sec:Pb}, we described the transition probability function $P$ of the stochastic game $\Gamma$ with respect to the base attack and reinforce success probabilities $\alpha^i_{o,s}$ and $\rho^i_{o,s}$ (for $(i,o,s) \in \{1,2\}\times\O\times \S$). Determining these probabilities requires soliciting input from military domain experts that could include wargamers, operational planners, and weapon system experts. In this case study, the base attack success probability $\alpha^i_{o,s}$ for Player $i$ on objective $o$ at a given state $s$ is calculated from two sets of inputs: an initial attack success probability $q^i_o$, and improvement probabilities $q^i_{o,O^\prime}$ for $O^\prime \subset \O$. Specifically, $q^i_{o,O^\prime}$ reduces the probability of the attack on $o$ failing when Player $i$ controls the set of objectives $O^\prime$. We construct the state-dependent base attack success probabilities as follows:
\begin{align}
\forall i \in \{1,2\}, \ \forall o \in \O, \ \forall s \in \S, \ \alpha^i_{o,s}=1-\left[(1-q^i_o)\cdot \hspace{0.5cm}\prod_{\mathclap{\{O^\prime \subset \O \, | \, s_{o^\prime} = i, \, \forall o^\prime \in O^\prime\}}} \hspace{0.5cm} (1-q^i_{o,O^\prime})\right].
\label{eq:pr}
\end{align}

The reinforce success probabilities are analogously constructed, and ensure that Assumptions \ref{ass:atk} and \ref{ass:rfc} are satisfied. In practice, each base attack and reinforce success probability is affected by a small number of subsets of objectives $O^\prime$.

We now illustrate the construction of Player 1's base attack success probability for \textit{NE Terrain} (obj. 9) given a state $s= (1,1,2,2,1,1,1,1,2,1,1,1,2,2,1,1,1,2,1,1,1,2)$, which corresponds to Player 1 controlling objectives $\{1,2,5,6,7,8,10,11,12,15,16,17,19,20,21\}$. The base attack success probability on objective 9 is given by $q^1_9=0.20$ and positive improvements due to the control of single objectives are given by $q^1_{9,\{12\}}=0.20$ and $q^1_{9,\{17\}}=0.10$. These probabilities reflect that the coalition could take \textit{NE Terrain} (obj. 9) with an initial probability of $0.20$; the control of \textit{Mountain Pass South} (obj. 12) closes the gap by $0.20$ and the control of \textit{Northern Waters} (obj. 17) closes the gap by $0.10$ due to naval fire support. We also consider the following boosts due to the control of multiple objectives: $q^1_{9,\{8,11,12\}}=0.15$ and $q^1_{9,\{7,11,16,20\}}=0.05$. These probabilities reflect the auxiliary force flow into \textit{NE Terrain} when Player 1 controls the \textit{Mountain Pass} (objs. 8 and 12 along with obj. 11 to ensure the Southern LoC is open to the pass) and the \textit{Canal} (objs. 7, 11, 16, 20). Given these five inputs, Player 1's resulting base attack success probability on objective 9 given this state $s$ is  $\alpha^1_{s,9}=1-(1-0.20)(1-0.20)(1-0.10)(1-0.15)(1-0.05) = 0.535$. This provides the probability of Player 1 controlling objective 9 in the subsequent state if Player 1 attacks objective 9 and Player 2 does \emph{not} reinforce it.

\subsection{Campaign Scenarios}

\begin{figure}[htbp]
\begin{subfigure}[h]{0.48\linewidth}
\includegraphics[width=\linewidth]{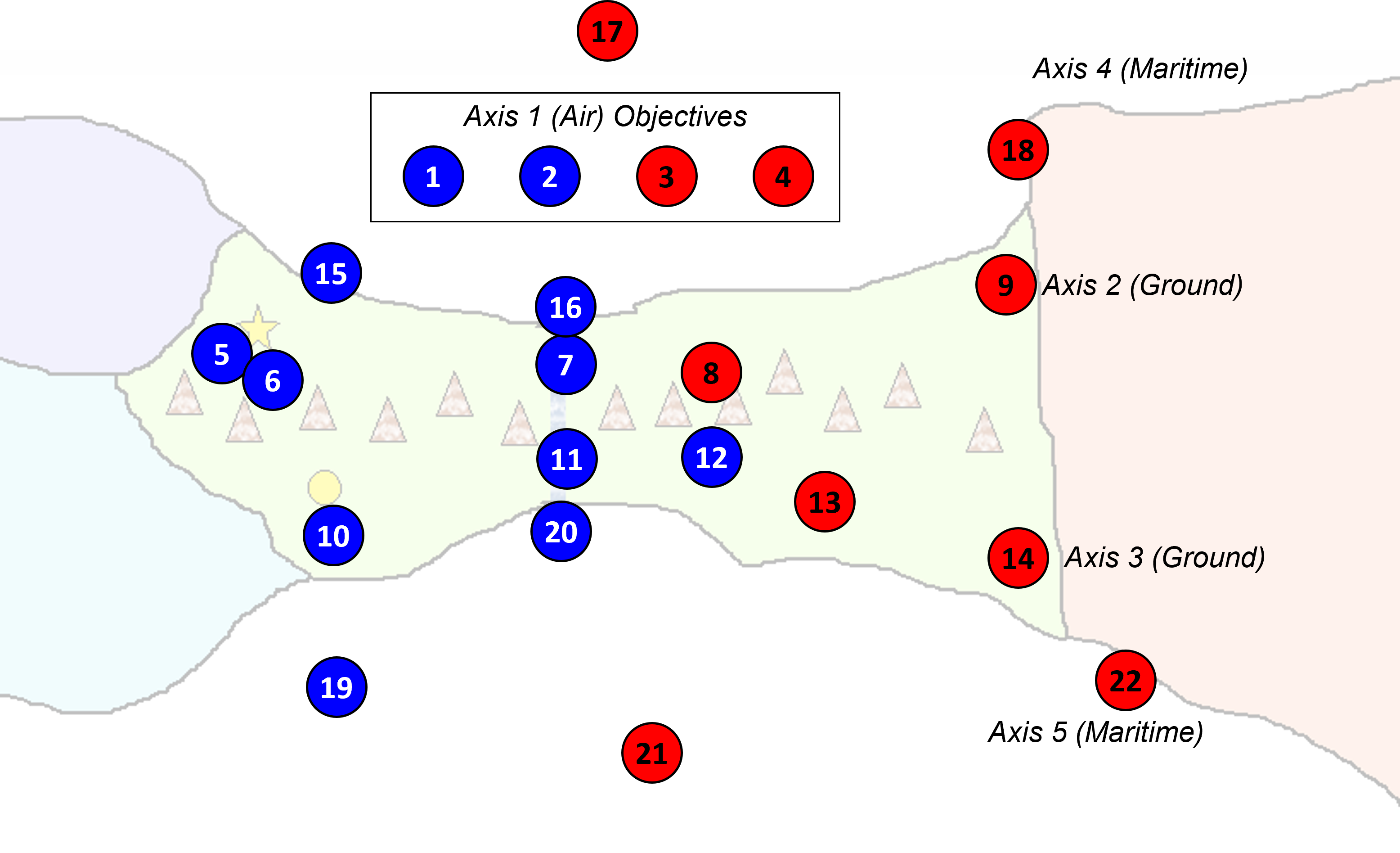}
\caption{Initial state $s^{1}$}
\end{subfigure}
\hfill
\begin{subfigure}[h]{0.48\linewidth}
\includegraphics[width=\linewidth]{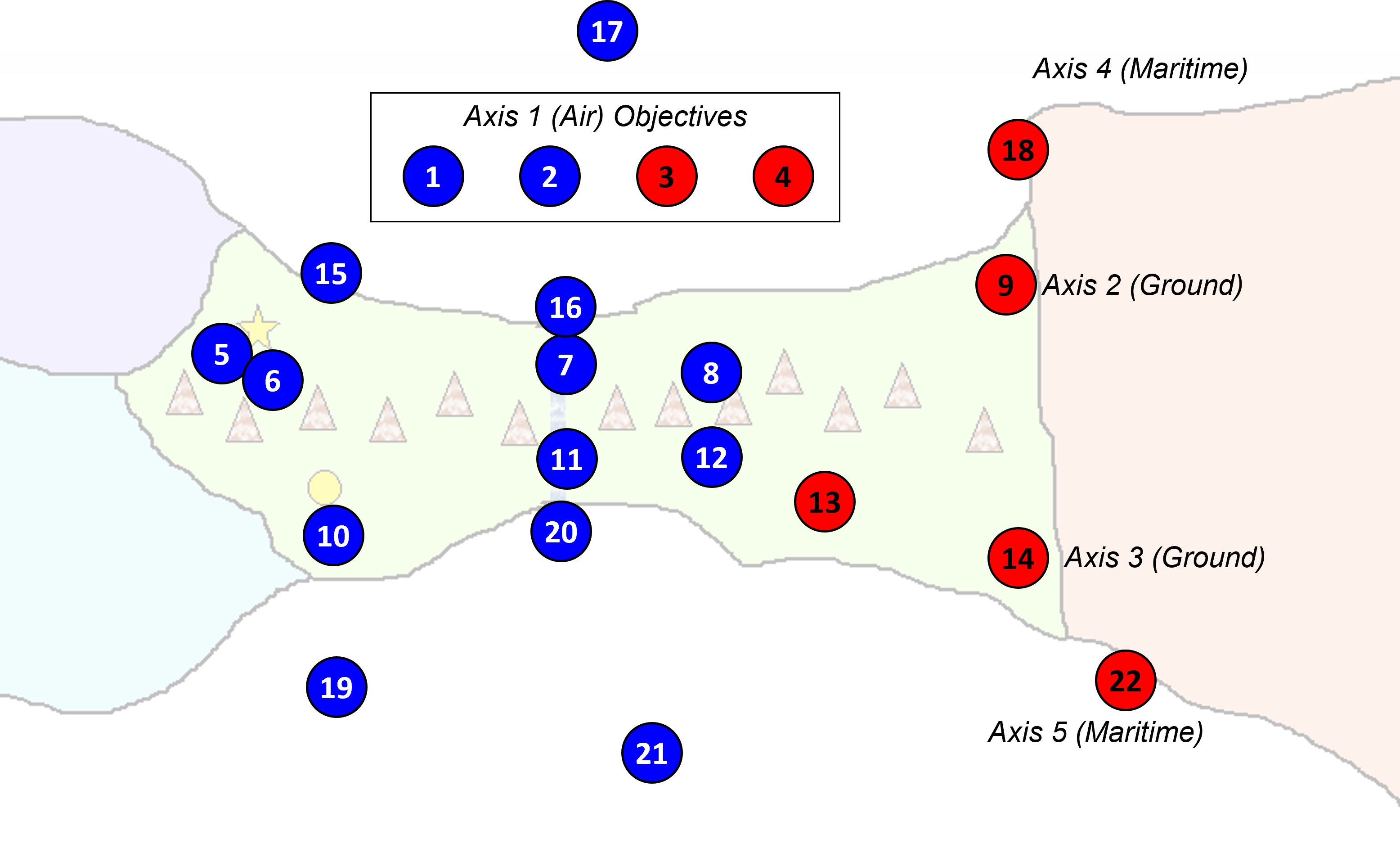}
\caption{Initial state $s^{2}$}
\end{subfigure}%
\caption{Campaign initial states. Objectives controlled by the coalition (resp. Adversary) are in blue (resp. red).}
\label{fig:fiction0}
\end{figure}

In this case study, we investigate and compare four campaign instances that differ in their initial states and transition probabilities. Specifically, we consider two initial campaign states, which are illustrated in Figure \ref{fig:fiction0}. We assume Adversary will control a subset of objectives at the campaign outset, so the initial state represents the campaign after their initial assault. In the first initial state, $s^{1}$, Adversary controls 10 of the 22 objectives at the campaign outset, while in the second initial state, $s^{2}$, Adversary controls 7 objectives.

Additionally, we assess a binary strategic decision: whether to invest in Isthia or not. We suppose the coalition has the possibility to invest in a 5th generation \emph{aircraft squadron} and \textit{military training} for Isthia's ground forces. The aircraft squadron decreases the probability of failure while attacking air objectives 2 and 3 by 10\%. The second component of the investment is a partnership program, which is an ongoing activity to train the Isthia infantry to defend their country. We suppose this will reduce Adversary's base attack success probabilities on objectives 8, 9, 12, and 13 by 20\%. This reflects the ability of the trained Isthia infantry to defend a subset of objectives without a commander giving a \textit{reinforce} order. 

\subsection{Impact of Initial State and Strategic Investment}

Given the two different initial states $s^1$ and $s^2$ and the binary strategic decision, we analyze four campaign scenarios. We solve each scenario using AVI, assuming a discount factor equal to $\gamma = 0.9$ between each stage, and an optimality gap given by $\epsilon = 0.001$. For the sake of comparison, we also compute the optimal value of the game for the initial state $s^A$ (resp. $s^I$) where Adversary (resp. Isthia) controls all objectives.
We present the corresponding optimal game values in Table \ref{tab:Values}.

\begin{table}[htbp]
\centering
\caption{Optimal game values for different initial states and investment decisions}
\label{tab:Values}
\begin{tabular}{r|c|c|c|c}
	\toprule
	  &  \multicolumn{4}{c}{Initial state}\\
	 \cline{2-5}
	& $s^A$ & $s^1$ & $s^2$ & $s^I$\\
 	 \midrule
	Without investment  & 59.22 & 21.92 & 11.48 & 3.75\\
        \hline
	With investment&  59.04 & 19.99 & \phantom{1}9.75  & 3.30  \\
	 \bottomrule
\end{tabular}
\end{table}

The first interesting observation is the wide differential in total expected loss across the initial states. Indeed, we find that without coalition investment, $V^*(s^1) - V^*(s^2) = 10.44$, which is significantly larger than the loss of objectives initially controlled by Adversary in $s^1$ but controlled by Isthia in $s^2$. Indeed, since Adversary controls more objectives in $s^1$ than in $s^2$, then $s^2 \preceq s^1$ and Theorem \ref{thm:isotone} guarantees that $V^*(s^1) - V^*(s^2) \geq \ell_8 + \ell_{17} + \ell_{21} = 0.8$. This shows that controlling more objectives at the campaign outset provides more value beyond the immediate reward (i.e., stage loss). This phenomenon is also exacerbated when comparing the two extremes, in which Adversary (resp. Isthia) initially controls all objectives in $s^A$ (resp. $s^I$).

Second, we find that the a priori strategic investment has a noticeable effect on the campaign, which varies with the initial state:  The investment results in an 8.8\% (resp. 15.1\%) reduction in optimal game value for state $s^1$ (resp. state $s^2$). The investment benefits the coalition from initial state $s^2$ more since Adversary is further inhibited from successfully attacking \textit{Mountain Pass North} (obj. 8). This shows the importance of properly timing an investment, as it will be more valuable for Isthia before Adversary controls too many objectives. 

Next, for each scenario, we examine the players' mixed strategies at the initial state in the  $\epsilon$-MPE obtained from the AVI algorithm. 
%Value iteration yields the strategy for 51,200 states. We present two states here. 
Tables \ref{tab:Mixed_1} and \ref{tab:Mixed_2} respectively compare Players 1 and 2's mixed strategies at state $s^1$ with and without investment.

\begin{table}[htbp]
\centering
% \begin{center}
\caption{Player 1's mixed strategy $\pi^1(s^1)$ in an $\epsilon$-MPE with $\epsilon=0.001$}
\label{tab:Mixed_1}
\begin{tabular}{c|c|c|c|c}
	\toprule
 \multicolumn{3}{c|}{Action $a^1$} & \multicolumn{2}{c}{Probability $\pi^1(s^1,a^1)$} \\\hline
	Commander 1& Commander 2& Commander 3 & Without investment & With investment  \\
 	\midrule
	\textit{atk} 3 & \textit{atk} \phantom{1}8 & \textit{atk} 21 & 0.986 & 0.876\\
        \hline
        \textit{atk} 3 & \textit{rfc} 12 & \textit{atk} 21 & 0.014 & 0\phantom{.014}\\
	\hline
        \textit{atk} 3 & \textit{rfc} \phantom{1}7 & \textit{atk} 17 & 0\phantom{.014} & 0.124\\
        \bottomrule
\end{tabular}
% \end{center}
\end{table}

\begin{table}[htbp]
\centering
\caption{Player 2's mixed strategy $\pi^2(s^1)$ in an $\epsilon$-MPE with $\epsilon=0.001$}
\label{tab:Mixed_2}
\begin{tabular}{c|c|c|c|c}
	\toprule
 \multicolumn{3}{c|}{Action $a^2$} & \multicolumn{2}{c}{Probability $\pi^2(s^1,a^2)$}  \\\hline
	Commander 1 & Commander 2& Commander 3 & Without investment & With investment \\
 	\midrule
	\textit{rfc} 3& \textit{atk} \phantom{1}7 & \textit{atk} 16 & 0.094 & 0.977\\
        \hline
        \textit{rfc} 3 & \textit{atk} 12 & \textit{atk} 16 & 0.906 & 0.023\\
        \bottomrule
\end{tabular}
\end{table}

We observe that whether the coalition invests or not, their air commander should always attack \textit{Airspace 2} (obj. 3) to attempt to gain air superiority, while Adversary reinforces that objective to thwart the coalition's attack.
Furthermore, when there is no investment, the coalition should always attack \textit{South Waters} (obj. 21). They are also interested in controlling the mountain pass to ensure free flow of supplies between the axes, which is why their ground commander attacks \textit{Mountain Pass North} (obj. 8) 98.6\% of the time and reinforces \textit{Mountain Pass South} (obj. 12) 1.4\% of the time (to reduce the likelihood of losing that objective to Adversary).
In contrast, Adversary always attacks \textit{Littoral North - Canal Access} (obj. 16), as they aim to provide naval fire support to their Northern ground forces. While waiting for naval support, Adversary's ground commander focuses their attack on \textit{Mountain Pass South} (obj. 12) 90.6\% of the time, and only attacks \textit{Canal North} (obj. 7) 9.4\% of the time.

However, the dynamics change when the coalition invests in an air squadron and training for Isthia ground forces. In particular, Isthia ground forces are now better trained to protect \emph{Mountain Pass South} (obj. 12), although the investment does not equip them to better protect \emph{Canal North} (obj. 7). As a result, Adversary's incentive is to refocus their ground effort, and instead attack \emph{Mountain Pass South} 2.3\% of the time, while attacking \emph{Canal North} 97.7\% of the time.

This has further consequences on Player 1's mixed strategy. Instead of reinforcing \emph{Mountain Pass South} (obj. 12) with some probability, their ground commander reinforces \emph{Canal North} (obj. 7) 12.4\% of the time. Interestingly, Player 1 coordinates their maritime and ground commanders, so that when the ground commander reinforces \emph{Canal North}, the maritime commander attacks \emph{Northern Waters} (obj. 17). This is primarily due to the interdependencies in the success probabilities between axes: While Player 1 maintains control of the whole \emph{Canal} (objs. 7, 11, 16, 20), the maritime commander receives more support to attack \emph{Northern Waters}. Hence, the investment both obviates the need to reinforce \emph{Mountain Pass South} due to the organic capabilities of the Isthian forces and introduces a new action which aligns the ground order to reinforce \emph{Canal North} with the naval order to attack \textit{Northern Waters}.

Now we present in Tables \ref{tab:Mixed_1.2} and \ref{tab:Mixed_2.2} the players' mixed strategies at the initial state $s^2$. While most insights are analogous to those for state $s^1$, we interestingly observe that at state $s^2$, Adversary always attacks \emph{Airspace 1} (obj. 2). Indeed, Adversary intended to maintain air parity at state $s^1$, and they are now interested in gaining air superiority at state $s^2$. This would provide them with air support to gain control of ground and maritime objectives, as more objectives are controlled by the coalition at $s^2$.

\begin{table}[htbp]
\begin{center}
\caption{Player 1's mixed strategy $\pi^1(s^2)$ in an $\epsilon$-MPE with $\epsilon=0.001$}
\label{tab:Mixed_1.2}
\begin{tabular}{c|c|c|c|c}
	\toprule
  \multicolumn{3}{c|}{Action $a^1$} & \multicolumn{2}{c}{Probability $\pi^1(s^2,a^1)$} \\\hline
	Commander 1 & Commander 2 & Commander 3 & Without investment & With investment  \\
 	\midrule
	\textit{atk} 3 & \textit{atk} 9 & \textit{atk} 22 & 0.677 & 0.681\\
        \hline
        \textit{atk} 3 & \textit{atk} 9 & \textit{rfc} 17 & 0.239 & 0.319 \\
	\hline
        \textit{atk} 3 & \textit{rfc} 8 & \textit{rfc} 17 & 0.084 & 0\phantom{.111}\\
        \bottomrule
\end{tabular}
\end{center}
\end{table}

\begin{table}[htbp]
\begin{center}
\caption{Player 2's mixed strategy $\pi^2(s^2)$ in an $\epsilon$-MPE with $\epsilon=0.001$}
\label{tab:Mixed_2.2}
\begin{tabular}{c|c|c|c|c}
	\toprule
  \multicolumn{3}{c|}{Action $a^2$} & \multicolumn{2}{c}{Probability $\pi^2(s^{(0)}=2,a^2)$} \\\hline
	Commander 1 & Commander 2 & Commander 3 & Without investment & With investment \\
 	\midrule
	\textit{atk} 2 & \textit{atk} \phantom{1}8 & \textit{atk} 17 & 0.552 & 0.561\\
        \hline
        \textit{atk} 2 & \textit{atk} \phantom{1}8 & \textit{atk} 21 & 0.424 & 0.439\\
	\hline
        \textit{atk} 2 & \textit{atk} 12 & \textit{atk} 21 & 0.024 & 0\phantom{.111}\\
        \bottomrule
\end{tabular}
\end{center}
\end{table}

We note that although Adversary desperately aims to gain control of \emph{Mountain Pass North} (obj. 8), Isthia barely reinforces this objective (or does not reinforce it at all with the coalition investment). This is primarily due to the support provided by the larger number of objectives in other axes controlled by the coalition, which impacts the game dynamics via two aspects. First, the coalition forces can defend reasonably well this objective without a reinforce order. Second, if the front is swapped---i.e., \emph{NE Terrain} (obj. 9) is gained by the coalition and \emph{Mountain Pass North} is gained by Adversary---then the coalition is likely to regain control of \emph{Mountain Pass North}, while maintaining control of \emph{NE Terrain} in the next turn. Finally, it is interesting to note that while Adversary aims to attack each Maritime front with similar probabilities, the coalition seeks to attack \emph{Adversary Seaport South} and fully control axis 5 or reinforce \emph{Northern Waters} and maintain the current disposition of axis 4.

A major takeaway from this analysis is that at each state in equilibrium, the mixed strategy for each player is extremely complex, as it depends on the state itself (i.e., who controls which objectives), the probabilistic interdependencies between axes, and the dynamics of the game (i.e., future states and actions). Furthermore, our examples demonstrate the need to carefully coordinate orders across commanders in order to achieve an equilibrium. Finally, we find that strategic investment decisions can significantly impact operational-level decision making for both players, and must be precisely analyzed. 

\subsection{Computational Performance}

Finally, we compare the performance of the solution algorithms described in Section \ref{sec:approaches} 
% on five campaigns of various sizes.
on the case study campaign (with 22 objectives, illustrated in Figure \ref{fig:22}) along with four other campaigns comprising 6, 10, 14, and 18 objectives, which are summarized in Table \ref{table:campaigns}. 

\begin{table}[htbp]
\begin{center}
\caption{Campaigns}
\begin{tabular}{c|c|c}
	\toprule
	\# objectives $|\O|$ & \# commanders $|\mathcal{C}|$ & \# axes $|\X|$\\ \midrule
	\phantom{1}6  & 1& 2\\ \hline 
        10 & 2  & 3  \\ \hline
        14 & 3 & 4  \\ \hline
        18 & 3 & 4  \\ \hline
        22 & 3 & 5  \\ \bottomrule
\end{tabular}
\label{table:campaigns}
\end{center}
\end{table}

For each campaign, we consider two data models. In the first model, we consider the entire original state space $\{1,2\}^\O$ and action spaces given by \eqref{actions_1}-\eqref{actions_2}. In the second model, we consider the reduced state and actions spaces given by \eqref{new:state} and \eqref{new_action_1}-\eqref{new_action_2}, as a consequence of Theorem \ref{thm:isotone} and Proposition \ref{prop:reduce}. Then, for each campaign and each data model, we run the VI and AVI algorithms with $\gamma = 0.9$ and $\epsilon= 0.001$. Table \ref{table:results} presents the results.
 
\begin{table}[htbp]
\begin{center}
\caption{Computational performance comparison}
\begin{tabular}{c|c|c|c|c|c|c} 
\toprule
	Data &  \multirow{2}{*}{Campaign} &  \multirow{2}{*}{\# states} & Max \# actions & \multicolumn{2}{c|}{Runtime [s]} & \# iterations\\ \cline{5-7}
 model & & & per state and player & VI & AVI & VI \& AVI\\\midrule
        % Full State, Full Action Shapley
        \multirow{5}{*}{$\begin{array}{c}\text{Original} \\ \text{state and action}\\ \text{spaces}\end{array}$}& \phantom{2}6 obj. & \phantom{4,194,3}64 & \phantom{43}6 & \phantom{20,0}12.5 & \phantom{8,00}1.1 & \phantom{1}96\\ \cline{2-7}
        & 10 obj. & \phantom{4,19}1,024 & \phantom{4}16 & \phantom{4}1,048.4 &  \phantom{8,0}74.4 & 102\\ \cline{2-7} 
        & 14 obj. & \phantom{4,1}16,384 & \phantom{4}72 & -- & -- & --\\ \cline{2-7} 
        & 18 obj. & \phantom{4,}262,144 & 160 &-- & -- & --\\ \cline{2-7} 
        & 22 obj. & 4,194,304 & 320 & -- & -- & --\\ \midrule
        % Full State, Reduced Action Shapley
	% \multirow{5}{2.5cm}{2: All States, Reduced Actions} & 6 obj & 64 & 16 & 6.5 sec. (97) & 0.8 sec. (97)   \\ \cline{2-6}
	%   & 10 obj & 1,024 & 64 & 216 sec. (104) & 23 sec. (104) \\ \cline{2-6}
 %        & 14 obj. & 16,384 & 256 &  6536 sec. (61) & 657 sec. (61) \\ \cline{2-6} 
 %        & 18 obj. & 262,144 & 256 & \multicolumn{2}{c|}{$>$ 24 hr.} \\ \cline{2-6}
 %        & 22 obj. & 4,194,304 &  1,024 & \multicolumn{2}{c|}{Not Completed} \\ \hline \hline
        % Structured Shapley
         \multirow{5}{*}{$\begin{array}{c}\text{Reduced} \\ \text{state and action}\\ \text{spaces}\end{array}$}& \phantom{2}6 obj & \phantom{4,194,3}64 & \phantom{43}4 & \phantom{20,00}2.8 &  \phantom{8,00}0.5  & \phantom{1}96\\ \cline{2-7}
	  & 10 obj & \phantom{4,194,}256 & \phantom{43}8 & \phantom{20,0}30.3&  \phantom{8,00}7.1 & 102\\ \cline{2-7}
        & 14 obj. & \phantom{4,19}2,304 & \phantom{4}16 & \phantom{20,}295.0 &  \phantom{8,0}85.7 & \phantom{1}61\\ \cline{2-7} 
        & 18 obj. & \phantom{4,19}6,400 &  \phantom{4}16 & \phantom{2}1,391.2&  \phantom{8,}397.0 & 102\\ \cline{2-7}
        & 22 obj. & \phantom{4,1}51,200 & \phantom{4}32& 21,310.2 & 8,654.3 & \phantom{1}64\\ \hline 
        % Structured Shapley
        %\multirow{5}{2.5cm}{3b: LoC Structure, CFH value initializes AVI} & 6 obj & 64 & $\leq$ 16 & 0.3 sec. \\ \cline{2-5}
	  %& 10 obj & 256 & $\leq$ 64 & 7.6 sec.\\ \cline{2-5}
        %& 14 obj. & 2,304 & $\leq$ 256 & 1.3 min.\\ \cline{2-5}
        %& 18 obj. & 6,400 & $\leq$ 256 & 5.8 min. \\ \cline{2-5}
        %& 22 obj. & 51,200 & $\leq$ 1,024 & 137.1 min. \\ \hline
        
\end{tabular}
\label{table:results}
\end{center}
\vspace{0.1cm}
\hfill \small -- : Runs out of memory
\end{table}

We first observe that the stochastic game $\Gamma$ with its original state and action spaces can only be solved for very small instances. For the larger campaigns, the computational challenge arises from the transition model $P$ exceeding the available memory and from having to iterate through the $2^{|\O|}$ states.

%The 14 obj. (and larger) campaigns with the original state and action spaces were not completed due to the size of the transition model: $P$ is too large for the available memory. We first observe that the stochastic game $\Gamma$ with its original state and action spaces can only be solved for very small instances. Even if $P$ may be stored, runtime is a challenge when iterating through the $2^{|\O|}$ states.

On the other hand, Table \ref{table:results} shows significant computational improvements when solving $\Gamma$ with its reduced state and action spaces. Indeed, thanks to the LoC and command structure, we were able to reduce the state space of the largest two instances by 97.5\% and 98.8\%, respectively. Furthermore, by showing isotonicity of the optimal value function (Theorem \ref{thm:isotone}) and properties of MPEs (Proposition \ref{prop:reduce}), we were able to further reduce the action space of each player (and at each state) by 90\% for these two instances. As a result, we find that the VI algorithm converges for all campaign instances.

Finally, our AVI algorithm empirically shows the value of reducing the number and/or size of linear programs to solve for this problem (as hinted by Proposition \ref{prop:reduction}). Even though it requires the same number of iterations as the classical VI algorithm, it completes each iteration faster, as it is often able to efficiently find pure equilibria of the matrix game, and is able to remove weakly dominated actions before solving the linear program. On average across all reduced state and action space campaigns, the AVI algorithm requires 28\% of the time (8\% of the time for the original state and action spaces) to converge compared to the classical VI algorithm. This numerical analysis demonstrates the computational value of leveraging the problem's structure and deriving equilibrium properties of the stochastic game $\Gamma$.

%!TEX root = NRL Manuscript.tex

\section{Conclusion} \label{sec:conclusion}

In this work, we studied what we believe is the first stochastic game model of a military campaign, extending the single-battle, single-stage matrix game model of \cite{haywood54}. We formulated a two-person discounted zero-sum stochastic game played by military forces over an infinite horizon. At each stage, each player orders each of their commanders to attack or reinforce an objective, which they can only execute if they have an open line of control from their base to that objective. When a battle occurs between the players, its outcome is stochastic and depends on the selected actions as well as the control of other objectives, accounting for the fire support between a player's forces. Each player aims to maximize the accumulated number of objectives they control, weighted by their criticality.

To solve this large-scale stochastic game and compute its Markov perfect equilibria, we first derived equilibrium properties of the game by leveraging its logistics and military operational command and control structure. In particular, we showed the nontrivial isotonicity of the optimal value function with respect to the partially ordered state space. This lead to a significant reduction of the state and action spaces, which in turn lead to the applicability of Shapley's value iteration algorithm. We further showed that the matrix game solved at each iteration of this algorithm admitted additional equilibrium properties, which we leveraged in our proposed accelerated value iteration that searches for pure equilibria or eliminates weakly dominated actions before solving the matrix game.

We then designed a case study with representative campaign instances to test our solution approach and derive new military insights enabled by the features of our stochastic game. Our analysis highlighted a complex interplay between the criticalities of the objectives, the probabilistic interdepedencies between objectives, and the dynamics of the game, leading to carefully designed equilibrium mixed strategies at every state. We also showed that strategic investment decisions can have a varied impact on operational-level decisions and performance, depending on the campaign outset, suggesting a careful timing of such investment decisions. Finally, our numerical analysis demonstrated the value of our equilibrium results in permitting us to efficiently solve the game with our accelerated value iteration algorithm for characteristic campaign instances that can be exploited by military decision makers.

A natural extension of this work is to consider a  geopolitical setting involving multiple adversaries, with possible heterogeneous discount factors to model differing player military sizes or importance of concluding the conflict. It would also be worthwhile to integrate strategic-level investment with operational-level planning, as motivated in our case study. Many countries must determine how to allocate resources between different forces located at different theaters, in anticipation of a conflict \citepalias{ar71-32}. Such extensions will lead to even larger and more complex games, requiring the development of faster approximation algorithms and heuristics, possibly using techniques from reinforcement learning.

\bibliography{references}
\bibliographystyle{authordate1}

\newpage
\renewcommand{\theHsection}{A\arabic{section}}

\begin{APPENDICES}

%!TEX root = NRL Manuscript.tex

\OneAndAHalfSpacedXI
\section{Proofs of Statements}\label{app:deriv}

%\md{In case it's needed at some point: The probability of success for player $i$ successfully reinforcing $o$ is expressed $(1-\rho^i_{o,s})$. We also know that $\alpha^i_{o,s} = 0 \text{ and } \rho^i_{o,s} = 1 \text{ for all } o\notin \overline{\O}^i_s$}

\proof{Proof of Theorem \ref{thm:isotone}.} Consider the sequence of functions $(V^{(t)})_{t \in \mathbb{Z}_{\geq0}}$ given by $V^{(0)}(s) = L(s)$ for every $s \in \S$ and $V^{(t+1)} = T(V^{(t)})$ for every $t \in \mathbb{Z}_{\geq0}$. We show by induction that:
\begin{align}
\forall t \in \mathbb{Z}_{\geq0}, \ \forall s \preceq s^\prime \in \S, \ V^{(t)}(s^\prime) - V^{(t)}(s) \geq \sum_{o \in \O} \ell_o  \cdot \mathds{1}_{\{s_o^\prime = 2 \text{ and } s_o = 1\}}. \label{eq:strict}
\end{align} 

By definition of the stage loss function,
\begin{align}
\forall s \preceq s^\prime \in \S, \ \ V^{(0)}(s^\prime) - V^{(0)}(s) = \sum_{o \in \O} \ell_o \cdot \mathds{1}_{\{s_o^\prime=2\}} - \sum_{o \in \O} \ell_o \cdot \mathds{1}_{\{s_o=2\}} =\sum_{o \in \O} \ell_o  \cdot \mathds{1}_{\{s_o^\prime = 2 \text{ and } s_o = 1\}}. \label{eq:init_V}
\end{align}

We now assume that \eqref{eq:strict} holds for a given $t \in \mathbb{Z}_{\geq0}$. We also consider two states $s^-\preceq s^+ \in \S$ that only differ in one given objective $o^\dag \in \O$, i.e., $s^-_{o^\dag} = 1$, $s^+_{o^\dag} = 2$, and $s^-_o = s^+_o$ for every $o \in \O\setminus\{o^\dag\}$. 

We define the mapping $\phi: \{\textit{atk,rfc,none}\}^{\O} \to \{\textit{atk,rfc,none}\}^{\O}$ that satisfies
%
%For every vector $a \in \{\textit{atk,rfc,none}\}^{\O}$, we define the following mapping:
\begin{align*}
\forall a \in \{\textit{atk,rfc,none}\}^{\O}, \ \forall o \in \O, \ \phi(a)_o =  \begin{cases}
\textit{rfc} & \text{ if } o = o^\dag \text{ and } a_o = \textit{atk}\\
a_o & \text{ otherwise}.
\end{cases}
\end{align*}

We note that for every action $a^{1} \in \A^1_{s^+}$, $\phi(a^1) \in  \A^1_{s^-}$. Similarly, for every $a^{2} \in \A^2_{s^-}$, $\phi(a^2) \in  \A^2_{s^+}$. We next show that
\begin{align*}
\forall a^{1} \in \A^1_{s^+}, \ \forall a^{2} \in \A^2_{s^-}, \ \mathbb{E}_{s^\prime \sim P(\cdot \, | \, s^-,\phi(a^{1}),a^{2})}[V^{(t)}(s^\prime)] \leq \mathbb{E}_{s^\prime \sim P(\cdot \, | \, s^+,a^{1},\phi(a^{2}))}[V^{(t)}(s^\prime)].
\end{align*}
Let  $a^{1} \in \A^1_{s^+}$, $a^{2} \in \A^2_{s^-}$, and $o \in \O$. If $o \neq o^\dag$, then we know that $s_o^- = s_o^+$, $\phi(a^{1})_o = a^{1}_o$ and $\phi(a^{2})_o = a^{2}_o$. Then, by Assumption \ref{ass:atk},  $\alpha_{o,s^-}^2 \leq \alpha_{o,s^+}^2$, $\rho_{o,s^-}^1 \geq \rho_{o,s^+}^1$, $\alpha_{o,s^-}^1 \geq \alpha_{o,s^+}^1$, and $\rho_{o,s^-}^2 \leq \rho_{o,s^+}^2$, which implies
\begin{align*}
p_o(1 \, | \, s^-,\phi(a^{1})_o,a^{2}_o) 
&= \begin{cases}
1 - \alpha^{2}_{o,s^-}\cdot\mathds{1}_{\{a^{2}_o = \textit{atk}\}} \cdot(1-\rho^1_{o,s^-}\cdot\mathds{1}_{\{\phi(a^{1})_{o} = \textit{rfc}\}})       &\text{if } s_o^-=1\\
\alpha^{1}_{o,s^-}\cdot\mathds{1}_{\{\phi(a^{1})_o = \textit{atk}\}} \cdot(1-\rho^2_{o,s^-}\cdot\mathds{1}_{\{a^{2}_{o} = \textit{rfc}\}})       &\text{if } s_o^-=2
\end{cases}\\
& \geq \begin{cases}
1 - \alpha^{2}_{o,s^+}\cdot\mathds{1}_{\{\phi(a^{2})_o = \textit{atk}\}} \cdot(1-\rho^1_{o,s^+}\cdot\mathds{1}_{\{a^{1}_{o} = \textit{rfc}\}})       &\text{if } s_o^+=1\\
\alpha^{1}_{o,s^+}\cdot\mathds{1}_{\{a^{1}_o = \textit{atk}\}} \cdot(1-\rho^2_{o,s^+}\cdot\mathds{1}_{\{\phi(a^{2})_{o} = \textit{rfc}\}})       &\text{if } s_o^+=2
\end{cases}\\
& = p_o(1 \, | \, s^+,a^{1}_o,\phi(a^{2})_o).
%&= \mathbb{P}(s_o^\prime = 1 \, | \, s^+,a^{1},\phi(a^{2})).
\end{align*}
%since by Assumption \ref{ass:atk},  $\alpha_{o,s^-}^2 \leq \alpha_{o,s^+}^2$ and $\rho_{o,s^-}^1 \geq \rho_{o,s^+}^1$ when $s_o^- = s_o^+ = 1$, and $\alpha_{o,s^-}^1 \geq \alpha_{o,s^+}^1$ and $\rho_{o,s^-}^2 \leq \rho_{o,s^+}^2$ when $s_o^- = s_o^+ = 2$.

For the remaining objective $o^\dag$, recall that $s_{o^\dag}^- = 1$ and $s_{o^\dag}^+ = 2$. If $a^{1}_{o^\dag} \neq  \textit{atk}$, then:
%\begin{align*}
%\mathbb{P}(s_{o^\dag}^\prime = 1 \, | \, s^-,\phi(a^{1}),a^{2}) \geq 0 = \mathbb{P}(s_{o^\dag}^\prime = 1 \, | \, s^+,a^{1},\phi(a^{2})).
%\end{align*}
\begin{align*}
p_{o^\dag}(1 \, | \, s^-,\phi(a^{1})_{o^\dag},a^{2}_{o^\dag}) \geq 0 = p_{o^\dag}(1 \, | \, s^+,a^{1}_{o^\dag},\phi(a^{2})_{o^\dag}).
\end{align*}

Similarly, if $a^{2}_{o^\dag} \neq  \textit{atk}$, then:
\begin{align*}
p_{o^\dag}(1 \, | \, s^-,\phi(a^{1})_{o^\dag},a^{2}_{o^\dag}) =1 \geq p_{o^\dag}(1 \, | \, s^+,a^{1}_{o^\dag},\phi(a^{2})_{o^\dag}).
\end{align*}

Finally, if $a^{1}_{o^\dag} = a^{2}_{o^\dag} = \textit{atk}$, then $\phi(a^{1})_{o^\dag} = \phi(a^{2})_{o^\dag} = \textit{rfc}$ and Assumption \ref{ass:rfc} implies
\begin{align*}
p_{o^\dag}(1 \, | \, s^-,\phi(a^{1})_{o^\dag},a^{2}_{o^\dag}) =1 - \alpha^{2}_{o^\dag,s^-}\cdot(1-\rho^1_{o^\dag,s^-}) \geq \alpha^{1}_{o^\dag,s^+}\cdot(1-\rho^2_{o^\dag,s^+})  = p_{o^\dag}(1 \, | \, s^+,a^{1}_{o^\dag},\phi(a^{2})_{o^\dag}).
\end{align*}

Thus, we obtain that
\begin{align*}
\forall a^{1} \in \A^1_{s^+}, \ \forall a^{2} \in \A^2_{s^-}, \ \forall o \in \O, \ p_o(1 \, | \, s^-,\phi(a^{1})_o,a^{2}_o)  \geq  p_o(1 \, | \, s^+,a^{1}_o,\phi(a^{2})_o),
\end{align*}
which implies that $p_o(\cdot \, | \,  s^+,a^{1}_o,\phi(a^{2})_o)$ first-order stochastically dominates $p_o(\cdot \, | \,s^-,\phi(a^{1})_o,a^{2}_o)$. Since $V^{(t)}$ is isotone, and by construction of the transition probability function, we then obtain the desired inequality:
%In other words, for every $a^{1} \in \A^1_{s^+}$, $a^{2} \in \A^2_{s^-}$, and $o \in \O$, 
%
%
%By inductive hypothesis, $V^{(t)}$ is isotone. Furthermore, by construction of the transition probability function, we obtain that
\begin{align*}
\forall a^{1} \in \A^1_{s^+}, \ \forall a^{2} \in \A^2_{s^-}, \ \mathbb{E}_{s^\prime \sim P(\cdot \, | \, s^-,\phi(a^{1}),a^{2})}[V^{(t)}(s^\prime)] \leq \mathbb{E}_{s^\prime \sim P(\cdot \, | \, s^+,a^{1},\phi(a^{2}))}[V^{(t)}(s^\prime)].
\end{align*}

A direct implication is that
\begin{align*}
\forall a^{1} \in \A^1_{s^+}, \ \forall a^{2} \in \A^2_{s^-}, \ &\ell_{o^\dag} + R(V^{(t)},s^-,\phi(a^{1}),a^{2}) =  L(s^+) + \gamma \cdot \mathbb{E}_{s^\prime \sim P(\cdot \, | \, s^-,\phi(a^{1}),a^{2})}[V^{(t)}(s^\prime)] \\
&\leq L(s^+) + \gamma \cdot\mathbb{E}_{s^\prime \sim P(\cdot \, | \, s^+,a^{1},\phi(a^{2}))}[V^{(t)}(s^\prime)] = R(V^{(t)},s^+,a^{1},\phi(a^{2})).
\end{align*}

Next, we define
\begin{align*}
\pi^{1^+}(s^+) \in \argmin_{\pi^1(s^+) \in \Delta(\A^1_{s^+})} \max_{a^2 \in \A^2_{s^+}} \mathbb{E}_{a^1 \sim \pi^1(s^+)}[R(V^{(t)},s^+,a^1,a^2)].
\end{align*}

In the final step, we construct $\pi^{1^-}(s^-) \in \Delta(\A^1_{s^-})$ defined by $\pi^{1^-}(s^-,\phi(a^{1})) = \pi^{1^+}(s^+,a^{1})$ for every $a^{1} \in \A^1_{s^+}$, which provides the desired result:
\begin{align*}
V^{(t+1)}(s^-) + \ell_{o^\dag} 
&\leq \max_{a^{2}\in \A^2_{s^-}} \sum_{a^{1}\in\A^1_{s^-}} \pi^{1^-}(s^-,a^{1}) \cdot (\ell_{o^\dag} + R(V^{(t)},s^-,a^{1},a^{2})) \\
&= \max_{a^{2}\in \A^2_{s^-}} \sum_{a^{1}\in\A^1_{s^+}} \pi^{1^-}(s^-,\phi(a^{1})) \cdot (\ell_{o^\dag} + R(V^{(t)},s^-,\phi(a^{1}),a^{2}))\\
& =  \max_{a^{2}\in \A^2_{s^-}} \sum_{a^{1}\in\A^1_{s^+}}\pi^{1^+}(s^+,a^{1}) \cdot (\ell_{o^\dag} + R(V^{(t)},s^-,\phi(a^{1}),a^{2}))\\
& \leq  \max_{a^{2}\in \A^2_{s^-}} \sum_{a^{1}\in\A^1_{s^+}}\pi^{1^+}(s^+,a^{1}) \cdot R(V^{(t)},s^+,a^{1},\phi(a^{2}))\\
& \leq  \max_{a^{2}\in \A^2_{s^+}} \sum_{a^{1}\in\A^1_{s^+}}\pi^{1^+}(s^+,a^{1}) \cdot R(V^{(t)},s^+,a^{1},a^{2})\\
& = V^{(t+1)}(s^+).
\end{align*}

By repeating this process, we deduce that for every $s\preceq s^\prime \in \S$, $V^{(t+1)}(s^\prime) - V^{(t+1)}(s) \geq \sum_{o \in \O} \ell_o \cdot \mathds{1}_{\{s^\prime_o = 2 \text{ and } s_o = 1\}}$. By induction, we then conclude that:
\begin{align*}
\forall t\in \mathbb{Z}_{\geq 0}, \ \forall s\preceq s^\prime \in \S, \ V^{(t)}(s^\prime) - V^{(t)}(s) \geq \sum_{o \in \O} \ell_o \cdot \mathds{1}_{\{s^\prime_o = 2 \text{ and } s_o = 1\}}.
\end{align*}

Taking the limit as $t$ goes to $+\infty$, we obtain:
\begin{align*}
\forall s\preceq s^\prime \in \S, \ V^{*}(s^\prime) - V^{*}(s) \geq  \sum_{o \in \O} \ell_o \cdot \mathds{1}_{\{s^\prime_o = 2 \text{ and } s_o = 1\}}.
\end{align*}

Thus, the optimal value of the game $\Gamma$ is an isotone function of the state space.
\hfill\Halmos
\endproof

\proof{Proof of Proposition \ref{prop:reduce}.}\

\textit{Property 1:} First, we show that under Assumption \ref{ass:init}, the set of achievable states for any policy profile is \eqref{new:state}. Consider a state $s \in \S$ and assume that $\type{x}{s} \in \{\textit{c1},\textit{c2},\textit{pf},\textit{sf}\}$ for every axis $x \in \X$. Let $a^1 \in \A^1_s$ and $a^2 \in \A^2_s$ be the actions drawn from the policies selected by both players and let $s^\prime$ be the subsequent state. Consider an individual axis $x = (o_1,\dots,o_n)$ of size $n \geq 2$ (since the case $n=1$ is trivial).

\begin{itemize}
\item[--] If $\type{x}{s} = \textit{c1}$ and $a^2_{o_n} = \textit{atk}$, then two possible outcomes occur: If the attack succeeds, then $s^\prime_{o_n} = 2$ and  $\type{x}{s^\prime} = \textit{pf}$ with the new front occurring at objectives $o_{n-1}$ and $o_{n}$. If instead the attack fails, then $s^\prime_{o_n} = 1$ and  $\type{x}{s^\prime} = \textit{c1}$. 

\item[--] If $\type{x}{s} = \textit{c2}$ and $a^1_{o_1} = \textit{atk}$, then two possible outcomes occur: If the attack succeeds, then $s^\prime_{o_1} = 1$ and  $\type{x}{s^\prime} = \textit{pf}$. If instead the attack fails, then $s^\prime_{o_1} = 2$ and  $\type{x}{s^\prime} = \textit{c2}$. 

\item[--] If $\type{x}{s} = \textit{pf}$ with the front occurring at $\{o_k,o_{k+1}\}$ for some $k \in \llbracket 1,n-1\rrbracket$, $a^1_{o_{k+1}} = \textit{atk}$, and $a^2_{o_{k}} \neq \textit{atk}$, then two similar outcomes may occur: If the attack succeeds, then $\type{x}{s^\prime} = \textit{pf}$ if $k< n-1$ or $\type{x}{s^\prime} =\textit{c1}$ if $k=n-1$. If the attack fails, then $\type{x}{s^\prime} = \textit{pf}$. Similar conclusions can be drawn if $a^2_{o_{k}} = \textit{atk}$, and $a^1_{o_{k+1}} \neq \textit{atk}$.

\item[--] If $\type{x}{s} = \textit{pf}$ with the front occurring at $\{o_k,o_{k+1}\}$ for some $k \in \llbracket 1,n-1\rrbracket$, $a^1_{o_{k+1}} = \textit{atk}$, and $a^2_{o_{k}} = \textit{atk}$, then additional outcomes may occur: If none or only one of the attacks succeeds, then $\type{x}{s^\prime} \in \{\textit{c1},\textit{c2},\textit{pf}\}$. If both attacks succeed, then $\type{x}{s^\prime} = \textit{sf}$ with a front for Player 1 at objective $\front{1}{x}{s^\prime} = \{o_{k}\}$ and another front for Player 2 at objective $\front{2}{x}{s^\prime} = \{o_{k+1}\}$.

\item[--] If $\type{x}{s} = \textit{sf}$ with a front for Player 1 at $\front{1}{x}{s^\prime} = \{o_{k}\}$ and another front for Player 2 at $\front{2}{x}{s^\prime} = \{o_{k+1}\}$ for some $k \in \llbracket 1,n-1\rrbracket$, then $a^1_{o_{k}} \in \{\textit{atk},\textit{none}\}$, $a^2_{o_{k+1}} \in \{\textit{atk},\textit{none}\}$, and multiple outcomes may occur: If one or two attacks are successful, then $\type{x}{s^\prime} \in \{\textit{c1},\textit{c2},\textit{pf}\}$. If no attack is successful, then $\type{x}{s^\prime} = \textit{sf}$.

%players are successful this will establish a \textit{pf} state. If the players are both unsuccessful \textit{sf} remains. Therefore, it is impossible to create further separation from a \textit{sf} state in the one-axis game. 

\item[--] In all other cases, $\type{x}{s^\prime} = \type{x}{s}$.
\end{itemize}

Thus, we obtain that $\type{x}{s^\prime} \in \{\textit{c1},\textit{c2},\textit{pf},\textit{sf}\}$ for every $x \in \X$. By induction, we conclude that at each stage of the campaign, each axis type will belong in $\{\textit{c1},\textit{c2},\textit{pf},\textit{sf}\}$. Henceforth, we assume that $\S$ is given by \eqref{new:state}.

\textit{Property 2:} Next, we show that in at least one equilibrium of the stochastic game $\Gamma$, the players randomize at every state over actions satisfying \eqref{new_action_1}-\eqref{new_action_2}.
%
%
%the action space of each player can be reduced to the actions satisfying \eqref{new_action_1}-\eqref{new_action_2}. 
%Consider an MPE $(\pi^{1^*},\pi^{2^*}) \in \Delta^1\times \Delta^2$ of $\Gamma$ and a state $s \in \S$. Then we know that $(\pi^{1^*}(s),\pi^{2^*}(s))$ is an equilibrium of the zero-sum matrix game $\Gamma(V^*,s)$. 
%
%
%
%We show that any of Player 1's best responses to $\pi^{2^*}(s)$ satisfies \eqref{new_action_1}-\eqref{new_action_2}. 
Consider a state $s \in \S$, a commander $c \in \C^1$, and let $a^{1^\dag} \in \A^1_s$ be an action for Player 1 such that $a^{1^\dag}_{o} = \textit{none}$ for every $o \in \cup_{x \in \X_c} \front{1}{x}{s}$. In other words, the commander $c$ for Player 1 does not reinforce nor attack any objective at any of the fronts in the axes under their responsibility. This is either because the commander $c$ reinforces an objective $o^\dag$ belonging to an axis $x^\dag$ that is not at the front (i.e., $a^{1^\dag}_{o^\dag} = \textit{rfc}$ with $o^\dag \in \overline{\O}^1_{s}\setminus \front{1}{x^\dag}{s}$), or commander $c$ does not attack nor reinforce any objective in that state. In the latter case, we still consider an arbitrary objective $o^\dag$ belonging to an axis $x^\dag \in \X_c$. 

Let us assume that there exists an axis $x^\prime \in \X_c$ such that 
 $\type{x^\prime}{s} \neq \textit{c1}$. Then, there exists $o^\prime \in \front{1}{x^\prime}{s}$ such that $s_{o^\prime} = 2$.
We next show that attacking $o^\prime$ provides a better or equal value for Player 1  in the zero-sum matrix game $\Gamma(V^*,s)$ regardless of Player 2's action. Let $a^{1^\prime}$ be defined as follows:
\begin{align}
\forall o \in \O, \ a^{1^\prime}_o = \begin{cases} \textit{atk} &\text{if } o = o^\prime \\ \textit{none} &\text{if } o = o^*\\ a^{1^\dag}_o & \text{otherwise}. \end{cases}\label{for_later 3}
\end{align}

Note that $a^{1^\prime} \in \A^1_s$. Then, for every $a_2 \in  \A^2_{s}$, we obtain the following:
\begin{align}
\forall o \in \O\setminus\{o^\prime\}, \ & p_o(1 \, | \, s,a^{1^\prime}_o,a^{2}_o) = p_o(1 \, | \, s,a^{1^\dag}_o,a^2_o)\nonumber\\
&p_{o^\prime}(1 \, | \, s,a^{1^\prime}_{o^\prime},a^{2}_{o^\prime}) = \alpha^1_{o^\prime,s}\cdot (1-\rho^2_{o^\prime,s} \cdot\mathds{1}_{\{a^2_{o^\prime} = \textit{rfc}\}}) \geq 0 = p_{o^\prime}(1 \, | \, s,a^{1^\dag}_{o^\prime},a^2_{o^\prime}).\label{for_later_1}
\end{align}

%And the inequality is strict for $a^{2^\prime}$ defined above.

Since $V^*$ is isotone (Theorem \ref{thm:isotone}), we deduce that:
%\begin{align*}
%\forall a^2 \in \A^2_{s}, \ \mathbb{E}_{s^\prime \sim P(\cdot\,|\,s,a^{1^\prime},a^2)}[V^*(s^\prime)] \leq \mathbb{E}_{s^\prime \sim P(\cdot\,|\,s,a^{1^\dag},a^2)}[V^*(s^\prime)],
%\end{align*}
%We finally obtain the following contradiction:
\begin{align}
\forall a^2 \in \A^2_{s}, \ R(V^*,s,a^{1^\prime},a^2) \leq R(V^*,s,a^{1^\dag},a^2).\label{for_later_2}
\end{align}

Thus, if $a^{1^\dag}$ is selected with positive probability in equilibrium, then we can construct another equilibrium that instead assigns that probability to $a^{1^\prime}$.

%Thus, either $a^{1^\dag}$ provides the same objective as  $a^{1^\prime}$ for any $a^2 \in \A^2_{s}$, or $a^{1^\dag}$ weakly dominates $a^{1^\prime}$

%Thus, $a^{1^\dag}$ is not a best response to $\pi^{2^*}(s)$ in the game $\Gamma(V^*,s)$ and cannot receive any positive probability in equilibrium.

%any  $a^{2^\dag} \in \A^2_s$ such that $a^{2^\dag}_{o} = \textit{none}$ for every $o \in \cup_{x \in \X_c} \front{2}{x}{s}$ is not a best response to $\pi^{1^*}(s)$ in $\Gamma(V^*,s)$ and cannot receive any positive probability in equilibrium.

%Consider a state $s \in \S$ and a commander $c \in \C$ for Player 1 such that 

Now, let us assume that for every axis $x \in \X_c$, $\type{x}{c} = \textit{c1}$. Let $x^\prime \in \X_c$ and $o^\prime \in \front{1}{x^\prime}{s}$.
%
%
%
%Since these axes are not of type \textit{c2}, the previous argument shows that $\pi^{2^*}(s)$ cannot assign any positive probability to any action $a^2 \in \A^2_s$ such that $a^{2}_{o} = \textit{none}$ for every $o \in \cup_{x \in \X_c} \front{2}{x}{s}$. Let $a^{2^\prime} \in \A^2_s$ be such that $\pi^{2^*}(s,a^{2^\prime}) > 0$. There exists an axis $x^\prime \in \X_c$ and an objective $o^\prime \in \front{1}{x^\prime}{s}$ such that $s_{o^\prime} = 1$ and $a^{2^\prime}_{o^\prime} = \textit{atk}$ (since commander $c$ for Player 2 cannot reinforce any objective under their responsibility when the axes are of type $\textit{c1}$). 
Then, we show that reinforcing $o^\prime$ provides a better or equal objective for Player 1   in $\Gamma(V^*,s)$ regardless of Player 2's action. Let $a^{1^\prime}$ be defined as follows:
\begin{align}
\forall o \in \O, \ a^{1^\prime}_o = \begin{cases} \textit{rfc} &\text{if } o = o^\prime \\ \textit{none} &\text{if } o = o^*\\ a^{1^\dag}_o & \text{otherwise}. \end{cases} \label{for_later4}
\end{align}

Note that $a^{1^\prime} \in \A^1_s$. Then, for every $a_2 \in  \A^2_{s}$, we obtain the following:
\begin{align}
&\forall o \in \O\setminus\{o^\prime\}, \ p_o(1 \, | \, s,a^{1^\prime}_o,a^{2}_o) = p_o(1 \, | \, s,a^{1^\dag}_o,a^2_o).\nonumber\\
&p_{o^\prime}(1 \, | \, s,a^{1^\prime}_{o^\prime},a^{2}_{o^\prime}) = 1 - \alpha^2_{o^\prime,s}\cdot \mathds{1}_{\{a^2_{o^\prime} = \textit{atk}\}}\cdot(1-\rho^1_{o^\prime,s}) \geq  1 - \alpha^2_{o^\prime,s}\cdot \mathds{1}_{\{a^2_{o^\prime} = \textit{atk}\}} = p_{o^\prime}(1 \, | \, s,a^{1^\dag}_{o^\prime},a^2_{o^\prime}).\label{for_later5}
\end{align}

%Furthermore, the previous inequality is strict for $a^{2^\prime}$ defined above. 
Since $V^*$ is isotone (Theorem \ref{thm:isotone}), we similarly deduce that:
%\begin{align*}
%\forall a^2 \in \A^2_{s}, \ \mathbb{E}_{s^\prime \sim P(\cdot\,|\,s,a^{1^\prime},a^2)}[V^*(s^\prime)] \leq \mathbb{E}_{s^\prime \sim P(\cdot\,|\,s,a^{1^\dag},a^2)}[V^*(s^\prime)],
%\end{align*}
%and the inequality is strict for $a^{2^\prime}$. Since $\pi^{2^*}(s,a^{2^\prime}) > 0$, we finally obtain the following contradiction:
\begin{align*}
\forall a^2 \in \A^2_{s}, \ R(V^*,s,a^{1^\prime},a^2) \leq R(V^*,s,a^{1^\dag},a^2).
\end{align*}
Thus, if $a^{1^\dag}$ is selected with positive probability in equilibrium, then we can construct another equilibrium that instead assigns that probability to $a^{1^\prime}$.

%Thus, $a^{1^\dag}$ is not a best response to $\pi^{2^*}(s)$ in the game $\Gamma(V^*,s)$ and cannot receive any positive probability in equilibrium.

A similar proof can be derived to show that there exists an equilibrium policy for Player 2 such that at every state $s \in \S$, no probability is assigned to an action $a^{2^\dag} \in \A^2_s$ satisfying $a^{2^\dag}_{o} = \textit{none}$ for every objective $o \in \cup_{x \in \X_c} \front{2}{x}{s}$ under the responsibility of some commander $c \in \C^2$.
Therefore, there exists an MPE of $\Gamma$ such that at any state $s\in \S$, each player randomizes over actions that satisfy \eqref{new_action_1}-\eqref{new_action_2}.

\textit{Property 3:} Finally, let us show that when the game parameters satisfy \eqref{cond1}-\eqref{cond3}, every action selected with positive probability in equilibrium at any state satisfies \eqref{new_action_1}-\eqref{new_action_2}. Let $s \in \S$ and assume that there exists a commander $c \in \C^1$ for Player 1 and an axis $x^\prime \in \X_c$ such that $\type{x^\prime}{s} \neq \textit{c1}$. Let $a^{1^\dag} \in \A^1_s$ be such that $a^{1^\dag}_{o} = \textit{none}$ for every $o \in \cup_{x \in \X_c} \front{1}{x}{s}$ and let $a^{1^\prime} \in \A^1_s$ defined in \eqref{for_later 3}. Since the game parameters satisfy \eqref{cond1}-\eqref{cond3}, then \eqref{for_later_1} becomes a strict inequality and $V^*$ becomes a strictly isotone function (Theorem \ref{thm:isotone}). Thus, \eqref{for_later_2} also becomes a strict inequality for every $a^2 \in \A^2_s$, and $a^{1^\dag}$ cannot receive any positive probability in equilibrium.

Similarly, if there exists a commander $c \in \C^2$ for Player 2 and an axis $x^\prime \in \X_c$ such that $\type{x^\prime}{s} \neq \textit{c2}$, then any  $a^{2^\dag} \in \A^2_s$ such that $a^{2^\dag}_{o} = \textit{none}$ for every $o \in \cup_{x \in \X_c} \front{2}{x}{s}$ cannot receive any positive probability in equilibrium.

Next, we consider a state $s \in \S$ and a commander $c \in \C^1$ for Player 1 such that for every axis $x \in \X_c$, $\type{x}{c} = \textit{c1}$. Let $x^\prime \in \X_c$ and $o^\prime \in \front{1}{x^\prime}{s}$ such that $s_{o^\prime} = 1$. Consider an MPE $(\pi^{1^*},\pi^{2^*}) \in \Delta^1\times \Delta^2$ of $\Gamma$. Since these axes are not of type \textit{c2}, the previous argument shows that $\pi^{2^*}(s)$ cannot assign any positive probability to any action $a^2 \in \A^2_s$ such that $a^{2}_{o} = \textit{none}$ for every $o \in \cup_{x \in \X_c} \front{2}{x}{s}$. Let $a^{2^\prime} \in \A^2_s$ be such that $\pi^{2^*}(s,a^{2^\prime}) > 0$. There exists an axis $x^\prime \in \X_c$ and an objective $o^\prime \in \front{1}{x^\prime}{s}$ such that $a^{2^\prime}_{o^\prime} = \textit{atk}$ (since commander $c$ for Player 2 cannot reinforce any objective under their responsibility when the axes are of type $\textit{c1}$). 

Let $a^{1^\prime} \in \A^1_s$ defined in \eqref{for_later4}. When the game parameters satisfy \eqref{cond1}-\eqref{cond3}, \eqref{for_later5} becomes a strict inequality for $a^{2^\prime}$ and $V^*$ a strictly isotone function (Theorem \ref{thm:isotone}). Since $\pi^{2^*}(s,a^{2^\prime}) > 0$, we deduce that:
\begin{align*}
\mathbb{E}_{a^2\sim \pi^{2^*}(s)} [R(V^*,s,a^{1^\prime},a^2)] < \mathbb{E}_{a^2\sim \pi^{2^*}(s)} [R(V^*,s,a^{1^\dag},a^2)],
\end{align*}
and $a^{1^\dag}$ cannot receive any positive probability in equilibrium.

A similar argument can be derived to show that if there exists a commander $c \in \C^2$ for Player 2 such that for every $x \in \X_c$, $\type{x}{s} = \textit{c2}$, then any $a^{2^\dag} \in \A^2_s$ such that $a^{2^\dag}_{o} = \textit{none}$ for every $o \in \cup_{x \in \X_c} \front{2}{x}{s}$ cannot receive any positive probability in equilibrium. In conclusion, when \eqref{cond1}-\eqref{cond3} hold, in \emph{every} MPE of $\Gamma$, each player randomizes at every state $s \in \S$ over actions that satisfy \eqref{new_action_1}-\eqref{new_action_2}.
\hfill\Halmos
\endproof

\proof{Proof of Proposition \ref{prop:reduction}.} Consider a campaign with one commander for each player, an iteration $t \in \mathbb{Z}_{>0}$ of the VI algorithm (initialized with $V^{(0)} = L$), and a state $s \in \S$. First, we investigate when the campaign has a single axis $x = (o_1,\dots,o_n)$ of size $n$.

% and a single axis $x = (o_1,\dots,o_n)$ of size $n$. Let $t \in \mathbb{Z}_{>0}$ be an iteration of the VI algorithm initialized with $V^{(0)} = L$, and $s \in \S$ be a state. 
 
% From the proof of Theorem \ref{thm:isotone}, we know that $V^{(t-1)}$ is an isotone function of the state space.

If $\type{x}{s} \in \{\textit{c1},\textit{c2},\textit{sf}\}$, then \eqref{new_action_1}-\eqref{new_action_2} imply that $|\A^1_s|= |\A^2_s| =1$ and $\Gamma(V^{(t-1)},s)$ admits a pure equilibrium. If $\type{x}{s} = \textit{pf}$ with  the front occurring at $\{o_k,o_{k+1}\}$ for some $k \in \llbracket 1,n-1\rrbracket$, let $s^-$ (resp. $s^+$) be the state that differs from $s$ at objective $o_{k+1}$ (resp. $o_{k}$). That is, $s^-_{o_{k+1}}=1$ and $s^+_{o_{k}}=2$.
%
% (i.e., $s^-_{o_{k+1}}=1$) and $s^+$ be the state that differs from $s$ at objective $o_{k}$ (i.e., $s^+_{o_{k}}=2$), 
%and $s^\prime$  be the state that differs from $s$ at both objectives $o_{k}$ and $o_{k+1}$ (i.e., $s^\prime_{o_{k}}=2$ and $s^\prime_{o_{k+1}}=1$). 
Each player $i \in \{1,2\}$ has two feasible actions at state $s$: the action $\hat{a}^i$ (resp. $\check{a}^i$) that attacks (resp. reinforces) the uncontrolled (resp. controlled) objective at the front. We obtain the following equalities:
\begin{align*}
&\mathbb{E}_{s^\prime \sim P(\cdot \, | \, s,\check{a}^1,\check{a}^2)}[V^{(t-1)}(s^\prime)] = V^{(t-1)}(s)\\
&\mathbb{E}_{s^\prime \sim P(\cdot \, | \, s,\hat{a}^1,\check{a}^2)}[V^{(t-1)}(s^\prime)] =p_{o_{k+1}}(1 \, | \, s,\hat{a}^1_{o_{k+1}},\check{a}^2_{o_{k+1}})\cdot V^{(t-1)}(s^-) + p_{o_{k+1}}(2 \, | \, s,\hat{a}^1_{o_{k+1}},\check{a}^2_{o_{k+1}})\cdot V^{(t-1)}(s)\\
&\mathbb{E}_{s^\prime \sim P(\cdot \, | \, s,\check{a}^1,\hat{a}^2)}[V^{(t-1)}(s^\prime)] = p_{o_{k}}(1 \, | \, s,\check{a}^1_{o_{k}},\hat{a}^2_{o_{k}})\cdot V^{(t-1)}(s) + p_{o_{k}}(2 \, | \, s,\check{a}^1_{o_{k}},\hat{a}^2_{o_{k}})\cdot V^{(t-1)}(s^+).
%R(V^{(t-1)},s,\hat{a}^1,\hat{a}^2) = &L(s) + \gamma \cdot \left(p_{o_{k}}(1 \, | \, s,\hat{a}^1_{o_{k}},\hat{a}^2_{o_{k}})\cdot p_{o_{k+1}}(1 \, | \, s,\hat{a}^1_{o_{k+1}},\hat{a}^2_{o_{k+1}}) \cdot V^{(t-1)}(s^-) \right.\\
%&+p_{o_{k}}(1 \, | \, s,\hat{a}^1_{o_{k}},\hat{a}^2_{o_{k}})\cdot p_{o_{k+1}}(2 \, | \, s,\hat{a}^1_{o_{k+1}},\hat{a}^2_{o_{k+1}}) \cdot V^{(t-1)}(s)\\
%&+p_{o_{k}}(2 \, | \, s,\hat{a}^1_{o_{k}},\hat{a}^2_{o_{k}})\cdot p_{o_{k+1}}(1 \, | \, s,\hat{a}^1_{o_{k+1}},\hat{a}^2_{o_{k+1}}) \cdot V^{(t-1)}(s^\prime)\\
%&\left.+p_{o_{k}}(2 \, | \, s,\hat{a}^1_{o_{k}},\hat{a}^2_{o_{k}})\cdot p_{o_{k+1}}(2 \, | \, s,\hat{a}^1_{o_{k+1}},\hat{a}^2_{o_{k+1}}) \cdot V^{(t-1)}(s^-).\right)
\end{align*}
%\begin{align*}
%R(V^{(t-1)},s,\check{a}^1,\check{a}^2) = &L(s) + \gamma \cdot V^{(t-1)}(s)\\
%R(V^{(t-1)},s,\hat{a}^1,\check{a}^2) = &L(s) + \gamma \cdot (p_{o_{k+1}}(1 \, | \, s,\hat{a}^1_{o_{k+1}},\check{a}^2_{o_{k+1}})\cdot V^{(t-1)}(s^-) + p_{o_{k+1}}(2 \, | \, s,\hat{a}^1_{o_{k+1}},\check{a}^2_{o_{k+1}})\cdot V^{(t-1)}(s) )\\
%R(V^{(t-1)},s,\check{a}^1,\hat{a}^2) = &L(s) + \gamma \cdot (p_{o_{k}}(1 \, | \, s,\check{a}^1_{o_{k}},\hat{a}^2_{o_{k}})\cdot V^{(t-1)}(s) + p_{o_{k}}(2 \, | \, s,\check{a}^1_{o_{k}},\hat{a}^2_{o_{k}})\cdot V^{(t-1)}(s^+) )\\
%%R(V^{(t-1)},s,\hat{a}^1,\hat{a}^2) = &L(s) + \gamma \cdot \left(p_{o_{k}}(1 \, | \, s,\hat{a}^1_{o_{k}},\hat{a}^2_{o_{k}})\cdot p_{o_{k+1}}(1 \, | \, s,\hat{a}^1_{o_{k+1}},\hat{a}^2_{o_{k+1}}) \cdot V^{(t-1)}(s^-) \right.\\
%%&+p_{o_{k}}(1 \, | \, s,\hat{a}^1_{o_{k}},\hat{a}^2_{o_{k}})\cdot p_{o_{k+1}}(2 \, | \, s,\hat{a}^1_{o_{k+1}},\hat{a}^2_{o_{k+1}}) \cdot V^{(t-1)}(s)\\
%%&+p_{o_{k}}(2 \, | \, s,\hat{a}^1_{o_{k}},\hat{a}^2_{o_{k}})\cdot p_{o_{k+1}}(1 \, | \, s,\hat{a}^1_{o_{k+1}},\hat{a}^2_{o_{k+1}}) \cdot V^{(t-1)}(s^\prime)\\
%%&\left.+p_{o_{k}}(2 \, | \, s,\hat{a}^1_{o_{k}},\hat{a}^2_{o_{k}})\cdot p_{o_{k+1}}(2 \, | \, s,\hat{a}^1_{o_{k+1}},\hat{a}^2_{o_{k+1}}) \cdot V^{(t-1)}(s^-).\right)
%\end{align*}

%\md{a few things can be removed}

Since $V^{(t-1)}$ is isotone, we deduce the following inequalities:
\begin{align*}
R(V^{(t-1)},s,\hat{a}^1,\check{a}^2)\leq R(V^{(t-1)},s,\check{a}^1,\check{a}^2) \leq R(V^{(t-1)},s,\check{a}^1,\hat{a}^2).
\end{align*}
%\begin{align*}
%& p_{o_{k+1}}(1 \, | \, s,\hat{a}^1_{o_{k+1}},\check{a}^2_{o_{k+1}})V^{(t-1)}(s^-) + p_{o_{k+1}}(2 \, | \, s,\hat{a}^1_{o_{k+1}},\check{a}^2_{o_{k+1}})V^{(t-1)}(s)  \leq V^{(t-1)}(s) \\
%&\leq p_{o_{k}}(1 \, | \, s,\check{a}^1_{o_{k}},\hat{a}^2_{o_{k}})V^{(t-1)}(s) + p_{o_{k}}(2 \, | \, s,\check{a}^1_{o_{k}},\hat{a}^2_{o_{k}})V^{(t-1)}(s^+).
%\end{align*}

Thus, a pure equilibrium of $\Gamma(V^{(t-1)},s)$ is
\begin{align*}
\begin{cases}
(\hat{a}^1,\check{a}^2) &\text{ if }  R(V^{(t-1)},s,\hat{a}^1,\hat{a}^2) \leq R(V^{(t-1)},s,\hat{a}^1,\check{a}^2)\\
(\hat{a}^1,\hat{a}^2) &\text{ if }  R(V^{(t-1)},s,\hat{a}^1,\hat{a}^2) \in [R(V^{(t-1)},s,\hat{a}^1,\check{a}^2), R(V^{(t-1)},s,\check{a}^1,\hat{a}^2)]\\
(\check{a}^1,\hat{a}^2) &\text{ if }  R(V^{(t-1)},s,\hat{a}^1,\hat{a}^2) \geq R(V^{(t-1)},s,\check{a}^1,\hat{a}^2).
\end{cases}
\end{align*}

Now, we investigate when the campaign has possibly multiple axes (under the responsibility of a single commander) and when the state $s$ is such that at least one axis $x = (o_1^x,\dots,o_{|x|}^x)$ is of type $\type{x}{s} = \textit{pf}$ with the front occurring at $\{o_k^x,o_{k+1}^x\}$ for some $k \in \llbracket 1,|x|-1\rrbracket$. We consider two feasible actions for each player $i \in \{1,2\}$: an action $\hat{a}^i$ (resp. $\check{a}^i$) that attacks (resp. reinforces) the uncontrolled (resp. controlled) objective at the front of axis $x$.

Consider $a^2 \in \A^2_s \setminus \{\hat{a}^2\}$. Then,
\begin{align*}
\forall o \in \O\setminus \{o_{k+1}^x\}, \ &p_o(1 \, | \, s, \hat{a}^1_o, a^2_o) = p_o(1 \, | \, s, \check{a}^1_o, a^2_o)\\
&p_{o_{k+1}^x}(1 \, | \, s, \hat{a}^1_{o_{k+1}^x}, a^2_{o_{k+1}^x}) \geq p_{o_{k+1}^x}(1 \, | \, s, \check{a}^1_{o_{k+1}^x}, a^2_{o_{k+1}^x}).
\end{align*} 
Since $V^{(t-1)}$ is isotone, we deduce that for every $a^2 \in \A^2_s\setminus \{\hat{a}^2\}$, $R(V^{(t-1)},s,\hat{a}^1,a^2) \leq R(V^{(t-1)},s,\check{a}^1,a^2)$. Thus, if the inequality $R(V^{(t-1)},s,\hat{a}^1,\hat{a}^2) \leq R(V^{(t-1)},s,\check{a}^1,\hat{a}^2)$ holds, then Player 1 will not reinforce $o_{k}^x$ in at least one equilibrium of $\Gamma(V^{(t-1)},s)$.

A similar argument shows that when the inequality $R(V^{(t-1)},s,\hat{a}^1,\hat{a}^2) \geq R(V^{(t-1)},s,\hat{a}^1,\check{a}^2)$ holds, then Player 2 will not reinforce $o_{k+1}^x$ in at least one equilibrium of $\Gamma(V^{(t-1)},s)$.

%A similar arguemtn
%
%
%Let us assume that $R(V^{(t-1)},s,\hat{a}^1,\hat{a}^2) \leq R(V^{(t-1)},s,\check{a}^1,\hat{a}^2)$ and c
%
%
%
%We consider two cases.
%
%Case 1: $R(V^{(t-1)},s,\hat{a}^1,\hat{a}^2) \leq R(V^{(t-1)},s,\check{a}^1,\hat{a}^2)$. Then we show that $\check{a}^1$ is weakly dominated by $\hat{a}^1$.
%
%Consider $a^2 \neq \hat{a}^2$, we obtain:
%\begin{align*}
%\forall o \neq o_{k+1}^x, \ p_o(1 \, | \, s, \hat{a}^1_o, a^2_o) = p_o(1 \, | \, s, \check{a}^1_o, a^2_o)\\
%p_{o_{k+1}^x}(1 \, | \, s, \hat{a}^1_{o_{k+1}^x}, a^2_{o_{k+1}^x}) \geq p_{o_{k+1}^x}(1 \, | \, s, \check{a}^1_{o_{k+1}^x}, a^2_{o_{k+1}^x}).
%\end{align*} 
%
%Since $V^{(t-1)}$ is isotone, we conclude that for every $a^2 \in \A^2_s$, $R(V^{(t-1)},s,\hat{a}^1,a^2) \leq R(V^{(t-1)},s,\check{a}^1,a^2)$.

Since $R(V^{(t-1)},s,\hat{a}^1,\check{a}^2) \leq R(V^{(t-1)},s,\check{a}^1,\check{a}^2) \leq R(V^{(t-1)},s,\check{a}^1,\hat{a}^2)$, then $R(V^{(t-1)},s,\hat{a}^1,\hat{a}^2)$ always satisfies at least one of the two above-mentioned inequalities, providing the desired result.
\hfill\Halmos
\endproof

\end{APPENDICES}
\end{document}